\documentclass[12pt]{JHEP3}
\usepackage{amsfonts} 
\usepackage{amssymb}
\usepackage{lscape}
\usepackage{cite}
\usepackage{epsfig}
\usepackage{multirow}
\usepackage{longtable}
\usepackage{amsmath}
\author{}

\newcommand{\be}{\begin{equation}}
\newcommand{\ee}{\end{equation}}
\newcommand{\ba}{\begin{array}}
\newcommand{\ea}{\end{array}}
\newcommand{\bea}{\begin{eqnarray}}
\newcommand{\eea}{\end{eqnarray}}

\newcommand{\ov}{\overline}
\def\IR{\relax{\rm I\kern-.18em R}}

\def\IP{\relax{\rm I\kern-.18em P}}
\def\inbar{\vrule height1.5ex width.4pt depth0pt}
\def\IC{\relax\,\hbox{$\inbar\kern-.3em{\rm C}$}}

\def\a{\alpha}
\def\b{\beta}
\def\c{\gamma}

\def\K3{{\bf K3}}

\def\b{\beta}

\def\ov{\overline}

\def\n2d{\cN_{V^*}^{\otimes 2}}

\def\IC{\mathbb{C}}

\def\IR{\mathbb{R}}

\def\IP{\mathbb{P}}

\def\cN{{\mathcal N}}

\def\nn{\nonumber}

\def\to{\rightarrow}

\title{Three- and Four-point correlators of excited bosonic twist fields}
\author{
Pascal Anastasopoulos$^{1}$\footnote{pascal@hep.itp.tuwien.ac.at},~
Mark D. Goodsell$^{2}$\footnote{mark.goodsell@cern.ch},~
Robert Richter$^{3}$\footnote{robert.richter@desy.de},~
\\
$^1$ Technische Univ. Wien Inst. f\"ur Theoretische Physik, A-1040 Vienna, Austria\\
$^2$ CPhT, Ecole Polytechnique, 91128 Palaiseau, France\\
$^3$ II. Institut f\"ur Theoretische Physik, Hamburg University, Germany\\
}

\date{}

\maketitle
\abstract{
We compute three- and four-point correlation functions containing excited bosonic twist fields. Our results can be used to determine properties, such as lifetimes and production rates, of massive string excitations localised at D-brane intersections, which could be signatures of a low string scale even if the usual string resonances  are inaccessible to the LHC.}
\preprint{
CPHT-RR049.0513\\
ZMP-HH/13-7\\
TUW-12-33}
 \clearpage
\begin{document}

\newpage

\section{Introduction}

Orientifold compactifications has been proven to be  a very fruitful framework for realistic model building. In those constructions the gauge symmetry does live on the world-volume of lower-dimensional hyperplanes, called D-branes, whereas the chiral matter is localised at the intersection of different D-brane stacks. For recent reviews on D-brane model building and concrete MSSM D-brane realizations, see \cite{Blumenhagen:2005mu, Blumenhagen:2006ci, Marchesano:2007de, Cvetic:2011vz} and references therein.

A particular intriguing subclass of D-brane realizations are string compactifications with a significantly lower string scale \cite{ArkaniHamed:1998rs, Antoniadis:1997zg, Antoniadis:1998ig}. Such low string scale scenarios with a string scale $M_s$ that can be as low as a few TeV provide a potential solution to the hierarchy and cosmological constant problems but may also lead to interesting signatures observable at the LHC. Among those signatures rank experimental signs of anomalous $Z'$ physics as well as Kaluza Klein states (see, e.g.\cite{Dudas:1999gz, Accomando:1999sj, Cullen:2000ef, Kiritsis:2002aj, Antoniadis:2002cs, Ghilencea:2002da, Anastasopoulos:2003aj, Anastasopoulos:2004ga,  Burgess:2004yq, Burikham:2004su, Chialva:2005gt, Coriano':2005js, Bianchi:2006nf, Anastasopoulos:2006cz, Anchordoqui:2007da,   Anastasopoulos:2008jt, Anchordoqui:2008ac, Lust:2008qc ,Anchordoqui:2008di, Armillis:2008vp, Fucito:2008ai, Anchordoqui:2009mm, Anchordoqui:2009ja, Lust:2009pz, Dong:2010jt, Feng:2010yx, Anchordoqui:2010zs,Cicoli:2011yy, Anchordoqui:2012wt, Anchordoqui:2011eg, Anchordoqui:2011ag}).

Low string scale compactifications also allow for the direct detection of string excitations. In a series of papers \cite{Lust:2008qc,Anchordoqui:2009mm,Lust:2009pz,Anchordoqui:2009ja} the authors study tree-level string scattering amplitudes containing at most two chiral fermions. They show that these amplitudes exhibit a universal behaviour independently of the specifics of the compactiﬁcation, which gives their results a predictive power. The observed poles correspond to the exchanges of Regge excitations of the standard model gauge bosons, whose mass scales like the string mass, and this has enabled the experiments to now limit the string scale to be above $5.08$ TeV \cite{ATLAS:2012qjz,ATLAS:2012pu, CMS:kxa}.

On the other hand at the intersection of two D-brane stacks exists a tower of stringy excitations, dubbed as \emph{light stringy states} in the following, whose mass scales as $M^2 \sim \theta M^2_s$, where $\theta$ denotes the intersection angle. Thus for small intersection angles those states can be significantly lighter than the first Regge excitations of the gauge bosons and are expected to be observed prior to the latter. Therefore it is of utmost interest to study properties, such as decay rates and life time, of these light stringy states.  

As recently discussed in \cite{Anastasopoulos:2011hj} the vertex operators of such light stringy states\footnote{Here the authors assumed as compactification manifold a factorizable six-torus $T^6=T^2 \times T^2 \times T^2$.} contain \emph{excited} versions of the bosonic twist fields present in the operators for massless chiral matter states (see e.g. \cite{Conlon:2011jq,Honecker:2011sm} for recent work involving these). However, little is known about correlators involving such excited bosonic twist fields\footnote{For some preliminary work on excited bosonic twist field correlators, see \cite{David:2000yn,Lust:2004cx}.}, a crucial ingredient in the computation of scattering amplitudes involving light stringy states.

This work is dedicated to filling this gap. For simplicity and reasons of calculability we assume the compactification manifold to be a factorizable six-torus $T^6=T^2 \times T^2 \times T^2$ wrapped by D6-branes. Such models are useful toys, but are still under active development (for original work, see e.g. \cite{Blumenhagen:2000wh,Angelantonj:2000hi,Aldazabal:2000cn,Aldazabal:2000dg,Forste:2000hx, Ibanez:2001nd, Cvetic:2001tj, Cvetic:2001nr,Honecker:2003vq, Cvetic:2004nk,Honecker:2004kb,Blumenhagen:2004xx,Blumenhagen:2005tn,Gmeiner:2005vz,Bailin:2006zf,Cvetic:2006by,Chen:2007px,Bailin:2007va,Bailin:2008xx,Gmeiner:2008xq},  for recent work, see e.g. \cite{Honecker:2012qr,Honecker:2012jd}). In that case scattering amplitudes split into three separate factorizable parts for each two-tori $T^2$, for which one can apply the developed CFT techniques. Using the latter we derive various three- and four-point correlators containing (higher) excited bosonic twist fields, where the latter live on one of three two-tori. More precisely we derive correlators of the type
 of type
\bea \nn
\langle \tau^+_{\alpha}  (x_1) \sigma^+_{\beta}  (x_2) \sigma^+_{\gamma} (x_3)\rangle\,\,\,  \\ \nn
\langle \tau^+_{\alpha} (x_1) \tau^+_{\beta}  (x_2) \sigma^+_{\gamma} (x_3) \rangle\,\,\, \\ \label{eq excited twist field correlator}
\langle \omega^+_{\alpha} (x_1) \sigma^+_{\beta}  (x_2) \sigma^+_{\gamma} (x_3) \rangle\,\,\,\\  \nn
\langle \tau^+_{\alpha} (x_1) \tau^-_{\alpha} (x_2) \sigma^+_{\beta}  (x_3) \sigma^-_{\beta} (x_4)   \rangle\,\,\, \\ \nn
\langle \omega^+_{\alpha} (x_1) \sigma^-_{\alpha} (x_2) \sigma^+_{\beta}  (x_3) \sigma^-_{\beta}  (x_4)   \rangle\,\,,
\eea
where the $\sigma$ fields are the usual bosonic twist fields while  $\tau$ and $\omega$ denote the excited and double excited bosonic twist fields. The subscript denotes the intersection angle which is measured in units of $\pi$ and ranges in the interval $[0,1)$.

Since the excited twist fields, $\tau$ and $\omega$ are not primary conformal fields we will take a detour to determine the correlators of type \eqref{eq excited twist field correlator} by first evaluating higher point correlators containing solely the primary fields, namely solely bosonic twist fields $\sigma$ as well as the conformal fields $\partial Z$ and $\partial \ov Z$. Given those correlators we obtain the correlators displayed in \eqref{eq excited twist field correlator} by applying various operator product expansions OPE's when performing appropriate limits of the higher point correlators.

This paper is organised as follows: In section \ref{sec general considerations} we illustrate the method which we use to evaluate the three and four-point correlators containing the excited bosonic twist fields. 
In section \ref{sec three-point}, we apply that method to compute the three-point correlation correlators while in section \ref{sec four-point} we derive the four-point correlators with one and two independent angles.
The appendices \ref{app OPE} and \ref{app integrals} contain necessary material for the derivation of the correlators. In appendix \ref{app vanishing} we demonstrate the vanishing of the quantum part of the correlators with only one excited field. In appendices \ref{app three point} and \ref{app four point} we display the remaining of our results which were omitted in the main text in order to keep the paper in a more readable/compact form.

\section{General considerations \label{sec general considerations}}

As already mentioned in the introduction we assume the compactification manifold to be a factorizable six-torus $T^6=T^2\times T^2 \times T^2$. In \cite{Cremades:2003qj,Cvetic:2003ch,Abel:2003vv,Abel:2003yx} \footnote{For similar work on the T-dual side with magnetized branes, see \cite{Cremades:2004wa}.} the authors determined the three and four-point bosonic twist field correlators in order to determine the physical Yukawa couplings as well as the K\"ahler metric for chiral matter \footnote{For a generalization to non-factorizable six-torus, see \cite{Bertolini:2005qh,Antoniadis:2009bg}.}. We will perform an analogous derivation for the three- and four-point correlators containing (higher) excited bosonic twist fields.

Let us illustrate the method which we will use to derive the above displayed correlators \eqref{eq excited twist field correlator}. 
Here we will focus on the three point correlators but the same procedure will apply to the four-point correlators. We start by determining the correlators containing three bosonic twist fields $\sigma$ and up to two of the bosonic conformal fields $\partial Z$ and $\partial \ov Z$. More explicitly we derive the four and five point correlators 
\begin{align}
\nn
\langle \partial Z(z)        \sigma^+_{\alpha} (x_1) \sigma^+_{\beta} (x_2) \sigma^+_{\gamma}(x_3) \rangle~~\\ \nn
\langle \partial \ov Z(z)  \sigma^+_{\alpha} (x_1) \sigma^+_{\beta}  (x_2) \sigma^+_{\gamma} (x_3) \rangle~~\\ \nn
\langle \partial Z(z) \partial  Z (w)  \sigma^+_{\alpha} (x_1) \sigma^+_{\beta}  (x_2) \sigma^+_{\gamma} (x_3) \rangle~~ \\ \label{eq five-point correlator}
\langle \partial  \ov Z(z) \partial \ov  Z (w) \sigma^+_{\alpha} (x_1) \sigma^+_{\beta}  (x_2) \sigma^+_{\gamma} (x_3)\rangle~~ \\ 
\langle \partial Z(z) \partial \ov  Z (w)  \sigma^+_{\alpha} (x_1) \sigma^+_{\beta} (x_2) \sigma^+_{\gamma} (x_3) \rangle  \nn\,\,.
\end{align}  

We have to keep in mind that the conformal fields $\partial Z$ and $\partial \ov Z$ do split into a classical piece $\partial Z_{cl}$ and $\partial \ov Z_{cl}$, respectively and a quantum piece denoted by $\partial Z_{qu}$ and $\partial \ov Z_{qu}$
\begin{align}
\partial Z= \partial Z_{cl} + \partial Z_{qu} \qquad \qquad \partial \ov Z= \partial \ov Z_{cl} + \partial \ov Z_{qu} \,\,.
\label{eq classical + quantum}
\end{align}
Here $\partial Z_{cl}$ and $\partial  \ov Z_{cl}$ satisfy the classical equation of motion with the boundary conditions set by the insertions of the bosonic twist fields. Thus the four- and five-point correlators split into a pure quantum part as well as a part which contains the classical piece, dubbed the \emph{mixed part} in what follows. 

For the derivation of the pure quantum correlator we will employ the energy momentum tensor method initially used in \cite{Hamidi:1986vh,Dixon:1986qv,Burwick:1990tu} in the context of closed string theory on orbifolds and more recently applied to open string theory \cite{Gava:1997jt,David:2000um,David:2000yn,Cvetic:2003ch,Abel:2003vv,Abel:2003yx}. On the other hand the mixed part requires the knowledge of the classical solutions for $\partial Z_{cl}$ and $\partial \ov Z_{cl}$ for the respective boundary conditions. Those were derived in \cite{Abel:2003vv,Abel:2003yx} for arbitrary number of bosonic twist field insertions as well as arbitrary intersection angles. Adopting their results for our specific setups in combination with known pure bosonic twist field three-point correlator \cite{Cvetic:2003ch, Lust:2004cx, Cvetic:2007ku} allows us to determine the mixed part.

Finally, any correlator is suppressed by the world-sheet instanton $e^{-S_{cl}}$, which is given by
\begin{align}
S_{cl}= \frac{1}{4 \pi  \alpha'} \int d^2 z \Big\{ \partial Z_{cl}(z) \partial Z_{cl}(z) +\partial \ov Z_{cl}(z) \partial \ov Z_{cl}(z) \Big\} \label{eq classical action}
\end{align}
with $\partial Z_{cl} $ and $\partial \ov Z_{cl}$ being again the classical solutions to the equation of motion satisfying the boundary conditions set by the bosonic twist fields. Given those solutions for the two-torus one can easily determine \eqref{eq classical action} where the whole world-sheet instanton contribution is a sum over all possible closed triangles connecting the three intersection points
\cite{Cremades:2003qj,Cvetic:2003ch, Abel:2003vv,Abel:2003yx} (See also \cite{Lust:2008qc})\footnote{For analogous results in the T-dual IIB framework with magnetized D-branes, see \cite{Cremades:2004wa,Antoniadis:2009bg}.}.

After establishing the complete four- and  five-point correlators \eqref{eq five-point correlator}, containing the pure quantum and the mixed part as well as the world-sheet contribution, we investigate various limits. Applying the OPE's
\begin{align} \nn
\partial Z (z) \,\sigma^+_{\alpha} (w)  &\sim (z-w)^{\alpha-1} \tau^+_{\alpha}(w)\qquad \qquad   \partial Z (z) \,\sigma^-_{\alpha} (w)  \sim (z-w)^{-\alpha} \tau^-_{\alpha}(w) \\
\partial \ov Z (z) \,\sigma^+_{\alpha} (w)  &\sim (z-w)^{-\alpha} \widetilde \tau^+_{\alpha}(w) \qquad \qquad \partial \ov Z (z) \,\sigma^-_{\alpha} (w)  \sim (z-w)^{\alpha-1} \widetilde \tau^-_{\alpha}(w) 
\label{eq OPE twist fields}
\end{align} 
will then give the three-point correlators containing the excited bosonic twist fields.

Even though, the procedure for the derivation of correlators containing exciting twist field correlators was laid out for three-point correlators it generalizes to four-point correlators as we will exemplify for four-point correlators  containing excited twist fields with one and two independent angles.  

\section{Three-point correlators\label{sec three-point}}
In this chapter we apply the previously described procedure to obtain three-point bosonic twist field correlators that contain excited bosonic twist fields. As laid out above we take a detour by first deriving higher point correlators containing the bosonic twist fields $\sigma_{\alpha}$, $\sigma_{\beta}$ and $\sigma_{\gamma}$ as well as the conformal fields $\partial Z$ and $\partial \ov Z$. The insertions of the bosonic twist fields $\sigma$ lead to the following boundary conditions
\begin{align}
\partial Z - \partial \ov Z =0  \qquad &\text{for} \qquad (-\infty, x_1)\\
e^{i \pi \alpha }\partial Z - e^{-i \pi \alpha }\partial \ov Z =0  \qquad &\text{for} \qquad (x_1, x_2)\\
e^{i \pi( \alpha +\beta) }\partial Z -e^{-i \pi( \alpha +\beta) } \partial \ov Z =0  \qquad &\text{for} \qquad  (x_2, x_3)\\
e^{i \pi( \alpha +\beta+ \gamma) }\partial Z -e^{-i \pi( \alpha +\beta+ \gamma) } \partial \ov Z =0  \qquad &\text{for} \qquad (x_3, \infty) \,\, .
\label{eq monodromy three-point}
\end{align}
In the following we assume that the sum of the intersection angles is $1$ rather than $2$, i.e. $\alpha +\beta +\gamma=1$. The results for the latter setup can be easily obtained making the replacement $\alpha, \beta, \gamma \rightarrow 1-\alpha, 1-\beta, 1-\gamma $.

\subsection{Three-point correlators with one excited twist field}
Let us start with  the three-point correlators containing one excited twist field. Those arise from the four-point correlators 
\begin{align} \nn
\langle \partial Z(z)        \sigma^+_{\alpha} (x_1) \sigma^+_{\beta} (x_2) \sigma^+_{\gamma} (x_3) \rangle\\ 
\langle \partial \ov Z(z) \sigma^+_{\alpha} (x_1) \sigma^+_{\beta} (x_2) \sigma^+_{\gamma} (x_3) \rangle \,\,,
\label{eq 4 point Z and ov Z}
\end{align}  
where we focus on a setup for which the intersection angles add up to $1$, i. e.  $\alpha+\beta+\gamma=1$. As indicated in section \ref{sec general considerations} the conformal fields $\partial Z$ as well as $\partial \ov Z$ split into a classical and a quantum piece, see eq. \eqref{eq classical + quantum}. As we show in appendix \ref{app vanishing} any bosonic twist field correlator containing an odd number of the conformal fields $\partial Z_{qu}$ and $\partial \ov Z_{qu}$ vanishes. Thus the four-point correlators \eqref{eq 4 point Z and ov Z} consist of a purely mixed part that takes the form
\begin{align} \nn
\partial Z_{cl}(z)    \, \langle         \sigma^+_{\alpha} (x_1) \sigma^+_{\beta} (x_2) \sigma^+_{\gamma} (x_3) \rangle~~ \\
\partial \ov Z_{cl} (z) \,\langle  \sigma^+_{\alpha} (x_1) \sigma^+_{\beta} (x_2) \sigma^+_{\gamma} (x_3)  \rangle\,\,.
\end{align} 
The three-point bosonic twist field correlator is given by \cite{Cvetic:2003ch, Lust:2004cx, Cvetic:2007ku}\footnote{In case the intersection angles add up to $2$ the correlator takes the form
\begin{align*}
\langle         \sigma^+_{\alpha} (x_1) \sigma^+_{\beta} (x_2) \sigma^+_{\gamma} (x_3) \rangle =\left( 2 \pi \frac{\Gamma(\alpha)\Gamma(\beta)\Gamma(\gamma)}{\Gamma(1-\alpha)\Gamma(1-\beta)\Gamma(1-\gamma)} \right)^{\frac{1}{4}} x^{-(1-\alpha) \,(1- \beta)}_{12}\, x^{-(1-\alpha) \, (1-\gamma)}_{13} \, x^{- (1-\beta) \,(1- \gamma)}_{23}\,\,.
\end{align*}
}
\begin{align}
\langle         \sigma^+_{\alpha} (x_1) \sigma^+_{\beta} (x_2) \sigma^+_{\gamma} (x_3) \rangle =\left( 2 \pi \frac{\Gamma(1-\alpha)\Gamma(1-\beta)\Gamma(1-\gamma)}{\Gamma(\alpha)\Gamma(\beta)\Gamma(\gamma)} \right)^{\frac{1}{4}} x^{-\alpha \, \beta}_{12}\, x^{-\alpha \, \gamma}_{13} \, x^{- \beta \, \gamma}_{23}\,\,.
\label{eq three point bosonic twist}
\end{align}
On the other hand the classical solutions are given by the  asymptotic behaviour close to the insertions of the bosonic twist fields $\sigma$, which specifies the solutions up to normalizations that can be evaluated from the global monodromy conditions. This procedure has been applied to three bosonic twist field correlators in \cite{Cremades:2003qj,Cvetic:2003ch, Abel:2003vv} \footnote{For analogous results in the T-dual IIB framework with magnetized D-branes, see \cite{Cremades:2004wa}.} where the authors derived the classical solutions for compactification manifold M, being planar $M= R^2$ and toroidal $M = T^2$. They take the form\footnote{Here we suppress the $x_{\infty}$ dependence and assume $z<0$.}
\begin{align}
\label{eq classical solution 3 point}
\partial Z_{cl} (z) &= e^{i \pi( \gamma-1) } z^{\alpha-1} (z-1)^{\beta-1} v_a  \frac{\Gamma(1-\gamma)}{\Gamma(\alpha) \,\Gamma(\beta)}\\ \nn
\partial \ov Z_{cl} (z) &=0\,\,,
\end{align}
where we used $SL(2,\mathbf{R})$ invariance to fix the bosonic twist field positions to $x_1=0$, $x_2=1$ and $x_3=\infty$. On a plane the vector $v_a$ denotes the distance between the two fixed points $v_a= f_1-f_2$, where $f_1$ and $f_2$ denote the intersection points of two D-brane stacks. However on a two-torus $T^2$ one has to take into account that the action of the twist operator does not only rotate $Z_{cl}$ but also may shift it by a lattice translation. That results into an infinite number of solutions with $v_a=f_1-f_2+ n \widetilde{L}_a$, where
\begin{align}
\widetilde{L}_a = \frac{|I_{bc}|}{gcd(|I_{ab}|, |I_{bc}|, |I_{ca})|} L_a\,\,.
\label{eq vector}
\end{align}
In eq. \eqref{eq vector}  $I_{xy} = n_x m_y -n_y m_x$ is the intersection number between two D-brane stacks $x$ and $y$, $L_x= R_1 \sqrt{n^2_x  + (m_x {\cal T})^2}$ the length of the D-brane stack $x$, with $(n_x, m_x)$ being the wrapping numbers and $ {\cal T}$ denoting the complex structure of the two-torus. In the simple setup in which all three D-brane stacks intersect each other exactly once $\widetilde{L}_x$ coincides with the length $L_x$ of the respective D-brane.

All correlators are suppressed by world-sheet instanton contributions which can be easily computed applying  \eqref{eq classical action} as well as using the classical solutions \eqref{eq classical solution 3 point}. One obtains
\begin{align}
S_{cl}= \frac{1}{ 2 \pi \alpha'}  \frac{\sin(\pi \alpha) \sin(\pi \beta)}{\sin(\pi \gamma)}  |v_a|^2\,\,,
\label{eq WS instanton three-point}
\end{align}
where $v_a$ is given by $v_a=f_1-f_2+ n \widetilde{L}_a$.
Thus the whole four-point correlator $\langle \partial Z      \sigma \, \sigma \, \sigma  \rangle$ takes the form \footnote{In the following we will suppress the $x_{\infty}$ factor, which we will recover when we reinstate the full $x_i$ dependence.}
\begin{align}
&\langle \partial Z(z)      \sigma^+_{\alpha} (0) \sigma^+_{\beta} (1) \sigma^+_{\gamma} (\infty) \rangle \\& \hspace{20mm}=
\sum_{n} v_a \left( 2 \pi \frac{\Gamma(1-\alpha)\Gamma(1-\beta)\Gamma^5(1-\gamma)}{\Gamma^5(\alpha)\Gamma^5(\beta)\Gamma(\gamma)} \right)^{\frac{1}{4}} e^{i \pi (\gamma-1)} z^{\alpha-1} (z-1)^{\beta-1} \,  e^{-S_{cl}} \nn
\label{eq Z sss}
\end{align}
with the world-sheet instanton given in eq. \eqref{eq WS instanton three-point} and the sum is over all closed triangles, thus over all integers $n$. On the other hand the four-point correlator $\langle \partial \ov Z      \sigma \, \sigma \, \sigma  \rangle$ vanishes.

Given the correlator \eqref{eq Z sss} we can now derive the three-point correlator containing an excited twist field. Performing the limit $z \rightarrow 0$ gives on 
\begin{align}
\langle      \tau^+_{\alpha} (0) \sigma^+_{\beta} (1) \sigma^+_{\gamma} (\infty) \rangle = \sum_{n}v_a \left( 2 \pi \frac{\Gamma(1-\alpha)\Gamma(1-\beta)\Gamma^5(1-\gamma)}{\Gamma^5(\alpha)\Gamma^5(\beta)\Gamma(\gamma)} \right)^{\frac{1}{4}}\,e^{-S_{cl}}\,\,,
\end{align}
where we used on the left-hand side the OPE's displayed in eq. \eqref{eq OPE twist fields}. Reinstating the complete $x_i$ dependence gives\footnote{Any 3-point corrleator of quasi-primary fields takes the form 
\begin{align*}
\langle \Phi_1 (x_1) \Phi_2 (x_2) \Phi_3 (x_3)  \rangle = \frac{C_{123}}{x^{h_1 + h_2 -h_3}_{12} \, x^{h_1 - h_2 +h_3}_{13} \, x^{-h_1 + h_2 +h_3}_{23} }\,\,,
\end{align*}
where $h_i$ denote the conformal dimension of the field $\Phi_i$.}
\begin{align}
\langle      \tau^+_{\alpha} (x_1) \sigma^+_{\beta} (x_2) \sigma^+_{\gamma} (x_3) \rangle = \sum_{n} v_a  \frac{ \left( 2 \pi \frac{\Gamma(1-\alpha)\Gamma(1-\beta)\Gamma^5(1-\gamma)}{\Gamma^5(\alpha)\Gamma^5(\beta)\Gamma(\gamma)} \right)^{\frac{1}{4}} }{ x^{\alpha(1+\beta)}_{12} x^{\alpha(1+\gamma)}_{13} x^{ \beta \gamma -\alpha}_{23} } \,\,e^{-S_{cl}} \,\,
\end{align}
with the world-sheet instanton contribution given by \eqref{eq WS instanton three-point}.

\subsection{Three-point correlators with two excited twist fields
\label{sec three-point two excited}}

Three-point correlators containing two excited bosonic twist field (or one double excited bosonic twist field) arise from the five point correlators of type
\begin{align}
\langle \partial Z(z) \partial \ov  Z (w)   \sigma^+_{\alpha}(x_1) \sigma^+_{\beta}(x_2) \sigma^+_{\gamma}(x_3)\rangle ~~ \nn\\
\langle \partial Z(z) \partial  Z (w)  \sigma^+_{\alpha}(x_1) \sigma^+_{\beta}(x_2) \sigma^+_{\gamma}(x_3) \rangle ~~\label{eq correlator 2 excited} \\
\langle \partial  \ov Z(z) \partial \ov  Z (w)\sigma^+_{\alpha}(x_1) \sigma^+_{\beta}(x_2) \sigma^+_{\gamma}(x_3) \rangle  \,\,. \nn
\end{align}
In contrast to the four point correlators discussed in the previous subsection the quantum parts of the correlators \eqref{eq correlator 2 excited} do not vanish. In order to determine those we define the following functions
\begin{align}
g(z,w) &= {\langle \partial Z_{qu}(z) \partial \ov Z_{qu}(w)\sigma^+_{\alpha}(x_1) \sigma^+_{\beta}(x_2) \sigma^+_{\gamma}(x_3) \rangle 
\over \langle \sigma^+_{\alpha}(x_1) \sigma^+_{\beta}(x_2) \sigma^+_{\gamma}(x_3)\rangle}\\
k(z,w) &= {\langle \partial  \ov Z_{qu}(z) \partial \ov Z_{qu}(w) \sigma^+_{\alpha}(x_1) \sigma^+_{\beta}(x_2) \sigma^+_{\gamma}(x_3)\rangle 
\over \langle \sigma^+_{\alpha}(x_1) \sigma^+_{\beta}(x_2) \sigma^+_{\gamma}(x_3)  \rangle}\\
m(z,w) &=  {\langle \partial Z_{qu}(z) \partial Z_{qu}(w) \sigma^+_{\alpha}(x_1) \sigma^+_{\beta}(x_2) \sigma^+_{\gamma}(x_3)\rangle 
\over \langle\sigma^+_{\alpha}(x_1) \sigma^+_{\beta}(x_2) \sigma^+_{\gamma}(x_3)\rangle}
\end{align} 
whose knowledge, together with the three-point correlator \eqref{eq three point bosonic twist}, allows us to derive the 5-point quantum correlator. With the local behaviour  \eqref{eq OPE twist fields} as well as 
\begin{align}
\partial Z (z) \partial \ov Z (w) \sim \frac{1}{(z-w)^2} \qquad \partial Z (z) \partial  Z (w) \sim \text{regular} \qquad\partial \ov Z (z) \partial \ov Z (w) \sim\text{regular}  
\label{eq OPE partial Z}
\end{align}
the form of the functions $g$, $k$ and $m$ is given by
\begin{align}
g(z,w) &=\omega_{1-\alpha,1-\beta,1-\gamma}(z) \omega_{\alpha, \beta, \gamma} (w) {P \over (z-w)^2}
\label{eq 3 ansatz g}\\
k(z,w) &= 0
\label{eq 3 ansatz k}\\
m(z,w) &=\omega_{1-\alpha,1-\beta,1-\gamma} (z) \omega_{1-\alpha,1-\beta,1-\gamma}  (w) C(\{x_i \})
\label{eq 3 ansatz m}
\end{align}
with
\begin{align}
\omega_{\alpha, \beta, \gamma} (z)=(z-x_1)^{-\alpha}(z-x_2)^{-\beta}(z-x_3)^{-\gamma}
\label{eq omega definition 3 point}
\end{align}
and
\begin{align}
P =& \alpha ~ (w-x_1)(z-x_2)(z-x_3)+ \beta ~ (z-x_1)(w-x_2)(z-x_3)\nn\\
&+\gamma ~ (z-x_1)(z-x_2)(w-x_3)\,\,.
\end{align}
Here we use the fact that any correlation function that contains a conformal field of dimension $h$ at insertion $x$ behaves as $x^{-2h}$ for $x \rightarrow \infty$.
While the functions $g$ and $k$ are fully determined by requiring the proper local behaviour for the full determination of $m$ we need to impose the monodromy conditions \eqref{eq monodromy three-point}, that results into the condition\footnote{The setup with three bosonic twist field insertions has only one independent world-sheet contour. Without loss of generality we choose the contour to be $0$ to $1$.}
\begin{align}
 e^{i\pi \alpha} \int^1_{0}   m(z,w) \, d w -  e^{-i\pi \alpha} \int^1_{0}  g(z,w)\, d w =0
\end{align}
which after dividing by $\omega_{1-\alpha,1-\beta,1-\gamma}(z)$, using $SL(2,\mathbf{R})$ symmetry to fix the bosonic twist field positions to
\begin{align}
x_1=0 \qquad x_2=1\qquad x_3=\infty 
\end{align} 
and performing the limit $z \rightarrow \infty$ results into 
\begin{align}
 e^{i\pi \alpha} \int^1_{0}   \widetilde m(w) \, d w -  e^{-i\pi \alpha} \int^1_{0} \widetilde{ g}(w)\, d w =0\,\,.
 \label{eq monodromy final 3 point}
\end{align}
Here $\widetilde g(w)$ and $\widetilde m(w)$ take the form
\begin{align}
\widetilde g(z)=   \gamma\, z^{-\alpha} (z-1)^{-\beta}  \qquad  \qquad 
\widetilde m(z)=  \widetilde{C} \,  z^{-1+\alpha} (z-1)^{-1+\beta}
 \end{align}
with $\widetilde{C}$ being $\frac{C(0,1,\infty)}{(-x_{\infty})^{2 \gamma-2}}$. Solving \eqref{eq monodromy final 3 point} allows to determine the missing piece of the function $m(z,w)$, $\widetilde{C}$, to
 \begin{align}
  \widetilde{C}=-e^{-2\pi i (\alpha+\beta)}\frac{\Gamma(1-\alpha)\Gamma(1-\beta)\Gamma(1-\gamma)}{\Gamma(\alpha)\Gamma(\beta)\Gamma(\gamma)}\,\,.
 \end{align}

Together with the three-point correlator \eqref{eq three point bosonic twist} we obtain the following five-point quantum correlators\footnote{Again we suppress the $x_{\infty}$ factors.}
\begin{align}
\langle \partial Z_{qu}(z) \partial \ov Z_{qu}(w)\sigma^+_{\alpha}(0) \sigma^+_{\beta}(1) \sigma^+_{\gamma}(\infty) \rangle &=\left( 2 \pi \frac{\Gamma(1-\alpha)\Gamma(1-\beta)\Gamma(1-\gamma)}{\Gamma(\alpha)\Gamma(\beta)\Gamma(\gamma)} \right)^{\frac{1}{4}} \\ & \hspace{-50mm} \times z^{\alpha-1} (z-1)^{\beta-1} w^{-\alpha} (w-1)^{-\beta} \bigg\{ \frac{\alpha w (z-1) + \beta z (w-1) + \gamma z (z-1)}{(z-w)^2} \bigg\} \nn \\
\langle \partial Z_{qu}(z) \partial Z_{qu}(w)\sigma^+_{\alpha}(0) \sigma^+_{\beta}(1) \sigma^+_{\gamma}(\infty) \rangle &=  - (2\pi)^{\frac{1}{4}}
\left(\frac{\Gamma(1-\alpha)\Gamma(1-\beta)\Gamma(1-\gamma)}{\Gamma(\alpha)\Gamma(\beta)\Gamma(\gamma)} \right)^{\frac{5}{4}}   \nn
\\ &\times  e^{2 \pi i \gamma}z^{\alpha-1} (z-1)^{\beta-1} w^{\alpha-1} (w-1)^{\beta-1}  \\
\langle \partial \ov Z_{qu}(z) \partial \ov Z_{qu}(w)\sigma^+_{\alpha}(0) \sigma^+_{\beta}(1) \sigma^+_{\gamma}(\infty) \rangle &=0 \,\,.
\end{align}
Since $\partial \ov Z_{cl}=0$ only $\langle \partial Z \partial Z \sigma^+_{\alpha} \sigma^+_{\beta} \sigma^+_{\gamma} \rangle $ gets contribution from the mixed part. Using \eqref{eq classical solution 3 point} the full five-point correlators take the form
\begin{align}
\label{eq 5 point correlator Z  ov Z}
\langle \partial Z(z) \partial \ov Z(w)\sigma^+_{\alpha}(0) \sigma^+_{\beta}(1) \sigma^+_{\gamma}(\infty) \rangle &=\left( 2 \pi \frac{\Gamma(1-\alpha)\Gamma(1-\beta)\Gamma(1-\gamma)}{\Gamma(\alpha)\Gamma(\beta)\Gamma(\gamma)} \right)^{\frac{1}{4}} \\ & \hspace{-50mm} \times z^{\alpha-1} (z-1)^{\beta-1} w^{-\alpha} (w-1)^{-\beta} \bigg\{ \frac{\alpha w (z-1) + \beta z (w-1) + \gamma z (z-1)}{(z-w)^2} \bigg\} \sum_n e^{-S_{cl}}\nn \\
\label{eq 5 point correlator Z Z}
\langle \partial Z(z) \partial Z(w)\sigma^+_{\alpha}(0) \sigma^+_{\beta}(1) \sigma^+_{\gamma}(\infty) \rangle &=-(2\pi)^{\frac{1}{4}}
\left(\frac{\Gamma(1-\alpha)\Gamma(1-\beta)\Gamma(1-\gamma)}{\Gamma(\alpha)\Gamma(\beta)\Gamma(\gamma)} \right)^{\frac{5}{4}}   
\\ & \hspace{-50mm} \times  e^{2 \pi i \gamma} z^{\alpha-1} (z-1)^{\beta-1} w^{\alpha-1} (w-1)^{\beta-1}  \sum_n \bigg\{ 1 - \frac{\sin(\pi \alpha) \sin(\pi \beta)}{\pi \sin(\pi \gamma)} |v_a|^2 \bigg\} \,\, e^{-S_{cl}} \nn \\
\langle \partial \ov Z(z) \partial \ov Z(w)\sigma^+_{\alpha}(0) \sigma^+_{\beta}(1) \sigma^+_{\gamma}(\infty) \rangle &=0 \,\,,
\label{eq 5 point correlator ov Z ov Z}
\end{align}
where the world-sheet instanton suppression is given in \eqref{eq WS instanton three-point} and the sum is over all closed triangles formed by the three D-brane intersections.

In order to derive the three-point correlators containing excited bosonic twist fields we take various limits of the five-point correlators, analogously to the derivation performed in the previous subsection. We illustrate this analysis with the example of the five-point correlator $\langle \partial Z \partial \ov Z\sigma^+_{\alpha} \sigma^+_{\beta} \sigma^+_{\gamma} \rangle$ and display the three-point correlators derived from the correlator $\langle \partial Z \partial  Z\sigma^+_{\alpha} \sigma^+_{\beta} \sigma^+_{\gamma} \rangle$ in appendix \ref{app three point}. 

From \eqref{eq 5 point correlator Z  ov Z} we obtain in the limit $z \rightarrow 0, w \rightarrow 1$ 
\begin{align}
\langle \tau^+_{\alpha}(0) \widetilde \tau^+_{\beta}(1) \sigma^+_{\gamma}(\infty) \rangle = \alpha \left( 2 \pi \frac{\Gamma(1-\alpha)\Gamma(1-\beta)\Gamma(1-\gamma)}{\Gamma(\alpha)\Gamma(\beta)\Gamma(\gamma)} \right)^{\frac{1}{4}} \,\, \sum_n  e^{-S_{cl}} \,\,.
\end{align} 
After reinstating the $x_i$ dependence one gets
\begin{align}
\langle \tau^+_{\alpha}(x_1) \widetilde \tau^+_{\beta}(x_2) \sigma^+_{\gamma}(x_3) \rangle =  \alpha \frac{\left( 2 \pi \frac{\Gamma(1-\alpha)\Gamma(1-\beta)\Gamma(1-\gamma)}{\Gamma(\alpha)\Gamma(\beta)\Gamma(\gamma)} \right)^{\frac{1}{4}}}{ x^{1+\alpha-(1-\alpha) \beta }_{12} x^{-(1-\alpha)\gamma}_{13} x^{(1+\beta)\gamma}_{23}} \,\, \sum_n e^{-S_{cl}} \,\,.
\end{align}
where again the world-sheet instanton suppression is given in \eqref{eq WS instanton three-point}.
Other three-point correlators can be derived in a similar fashion from \eqref{eq 5 point correlator Z Z} and we display them in appendix \ref{app three point}.

\section{Four-point correlators}\label{sec four-point}  

In this chapter we apply the same formalism as above to determine four-point correlators containing excited bosonic twist fields. We will discuss two different scenarios, namely bosonic twist field four-point correlators with one and two independent angles, starting with the case of one independent angle.

\subsection{Four-point correlators with one independent angle
\label{sec 4-point one angle}}
As for the three-point correlators we derive the four-point correlators containing by computing higher point correlators containing the bosonic twist fields $\sigma^+_{\alpha}$ and $\sigma^-_{\alpha}$ as well as $\partial Z$ and $\partial \ov Z$. The insertion of the bosonic twist fields $\langle \sigma^+_{\alpha} (x_1) \sigma^-_{\alpha} (x_2) \sigma^+_{\alpha} (x_3) \sigma^-_{\alpha} (x_4)\rangle $ lead to the following boundary conditions on the world-sheet
\begin{align}
\partial Z -  \partial \ov Z = 0& \,\,\,\,\, \text{for} \,\,\,\,(-\infty, x_1) \cup (x_2,x_3) \cup  (x_4, \infty)\nn\\
e^{i \pi \alpha}\partial Z - e^{-i\pi \alpha} \partial \ov Z = 0&  \,\,\,\,\, \text{for} \,\,\,\,(x_1,x_2) \cup (x_3,x_4) \,\,.
\label{eq boundary conditions}
\end{align}
Given those boundary conditions we will determine again the correlators containing one or two excited bosonic twist fields, starting with the correlators containing only one excited bosonic twist field.

\subsubsection{Four-point correlators containing one excited twist field \label{sec four-point one exc one angle}}
Analogously to the three-point correlator derivation we determine the four-point correlators containing one excited bosonic twist field by first computing the 5-point correlators 
\begin{align}
\langle \partial Z(z)  \sigma^+_{\alpha} (x_1) \sigma^-_{\alpha} (x_2) \sigma^+_{\alpha} (x_3) \sigma^-_{\alpha} (x_4) \rangle   \label{eq 5 point dZ correlator 1}~~\\
\langle \partial  \ov Z(z) \sigma^+_{\alpha} (x_1) \sigma^-_{\alpha} (x_2) \sigma^+_{\alpha} (x_3) \sigma^-_{\alpha} (x_4)  \rangle \,\,,
\label{eq 5 point dZ correlator}
\end{align} 
which get only a contribution from the classical part of $\partial Z $ and $\partial \ov Z$, respectively.  

Again the asymptotic behaviour close to the insertion of the bosonic twist fields specify the classical solutions $\partial Z_{cl}$ and $\partial \ov Z_{cl}$ up to normalizations
\begin{align} 
\partial Z_{cl} (z) =  \widetilde{a} ~\omega_{1-\alpha,1-\alpha}(z) \qquad \qquad \partial \ov Z_{cl} (z) = ~\widetilde{b}~ \omega_{\alpha,\alpha}(z) 
\label{eq classical sol one independent}
\end{align}
with $\omega_{\alpha,\beta}$ being
\begin{align}
\omega_{\alpha,\beta} (z)
=(z-x_1)^{-\alpha}(z-x_2)^{-1+\alpha}(z-x_3)^{-\beta}(z-x_4)^{-1+\beta}\,\,.
\label{eq omega definition}
\end{align}
The normalizations $\widetilde{a}$ and $\widetilde{b}$ are determined by satisfying the global monodromy conditions. One obtains \cite{Cvetic:2003ch, Abel:2003yx}\footnote{Here we used $SL(2,\mathbf{R})$ invariance to fix the twist field insertions to $x_1=0$, $x_2=x$, $x_3=1$ and $x_4=x_{\infty}=\infty$. Furthermore we suppress all $x_{\infty}$ dependence.} 
\begin{align}
\widetilde{a} &=- {\sin(\pi \alpha)\over 2 \pi} {v_b+e^{i\pi \alpha} v_a \tau(x)\over _2F_1[\alpha,1-\alpha,1,1-x]}\nn\\
\widetilde{b} &=- {\sin(\pi \alpha)\over 2 \pi} {v_b-e^{i\pi \alpha} v_a \tau(x) \over _2F_1[\alpha,1-\alpha,1,1-x]}\,\,,
\end{align}
where $\tau(x)$ is given by the ratio
\begin{align}
\tau(x)=\frac{_2F_1[\alpha,1-\alpha,1,1-x]}{_2F_1[\alpha,1-\alpha,1,x]}
\label{eq definition of tau 1}
\end{align}
and $v_a=f_2-f_1 + p \widetilde{L}_a $, $v_b= f_3-f_2 + q \widetilde{L}_b$ with $p$, $q$ being integers and $\widetilde{L}_x$ given in eq. \eqref{eq vector}. 
Together with the four point correlator 
\cite{Gava:1997jt,David:2000um,David:2000yn,Cvetic:2003ch,Abel:2003vv,Abel:2003yx,Anastasopoulos:2011gn}
\begin{align}
\label{eq 4 sigmas 1 angle} 
&\langle \sigma^+_{\alpha}(0) \sigma^-_{\alpha}(x) \sigma^+_{\alpha}(1) \sigma^-_{\alpha}(\infty) \rangle\\ & \hspace{20mm}={\frac{ \sin^\frac{1}{2}(\pi \alpha)}{ _2F_1[\alpha,1-\alpha,1,x]^\frac{1}{2}\,    _2F_1[\alpha,1-\alpha,1,1-x]^\frac{1}{2}} } \left[x (1-x) \right]^{-\alpha(1-\alpha)}  \nn
\end{align}
one obtains for the 5-point correlators \eqref{eq 5 point dZ correlator 1} and \eqref{eq 5 point dZ correlator} 
\begin{align} \nn
\langle \partial Z(z)  \sigma^+_{\alpha} (0) \sigma^-_{\alpha} (x) \sigma^+_{\alpha} (1) \sigma^-_{\alpha} (\infty) \rangle &=  - e^{-i \pi \alpha}z^{\alpha-1} (z-x)^{-\alpha} (z-1)^{\alpha-1} \left[x (1-x) \right]^{-\alpha(1-\alpha)}
\\ & \hspace{-50mm}\times
{\frac{ \sin^\frac{3}{2}(\pi \alpha)}{2 \pi\,  _2F_1[\alpha,1-\alpha,1,x]^\frac{1}{2}\,    _2F_1[\alpha,1-\alpha,1,1-x]^\frac{3}{2}} } 
\sum_{p,q}\left( {v_b+e^{i\pi \alpha} v_a  \tau(x)} \right)
 \,e^{-S_{cl}} \\ \nn
\langle \partial  \ov Z(z) \sigma^+_{\alpha} (0) \sigma^-_{\alpha} (x) \sigma^+_{\alpha} (1) \sigma^-_{\alpha} (\infty)  \rangle & =
- e^{i \pi (\alpha-1)} z^{-\alpha} (z-x)^{\alpha-1} (z-1)^{-\alpha} \left[x (1-x) \right]^{-\alpha(1-\alpha)}
\\ &\hspace{-50mm}\times
{\frac{ \sin^\frac{3}{2}(\pi \alpha)}{2 \pi\,  _2F_1[\alpha,1-\alpha,1,x]^\frac{1}{2}\,    _2F_1[\alpha,1-\alpha,1,1-x]^\frac{3}{2}} } 
\sum_{p,q} \left( {v_b-e^{i\pi \alpha} v_a  \tau(x)} \right)
\,e^{-S_{cl}} \,\,.
\label{eq dZ s s s s one angle}
\end{align}
 As already indicated both correlators are suppressed by the world-sheet instanton contributions $e^{-S_{cl}}$ which we discuss momentarily.
 With \eqref{eq classical action} and \eqref{eq classical sol one independent} we can compute the classical action in the presence of the four bosonic twist fields compute to\cite{Abel:2003yx} (see also \cite{Lust:2008qc})
\begin{align}
S_{cl}^{T^2}= \frac{\pi}{\alpha'} \sin(\pi a) \Big\{ |v_a|^2 \tau(x) + |v_b|^2 \tau(1-x)  \Big\} \,\,,
\label{eq WS torus one angle}
\end{align}
where $\tau(x)$ is given by \eqref{eq definition of tau 1}.
The complete world-sheet instanton contribution is the sum over all possible closed polygons connecting the four intersection points, thus we sum over all integers $p$ and $q$ appearing in $v_a$ and $v_b$.

In the following we will perform the limits $z \rightarrow 0$ and $z\rightarrow x$ in order to derive the four point correlators with one excited bosonic twist field. We will illustrate that with the example $\langle  \partial Z  \sigma^+ \sigma^- \sigma^+ \sigma^- \rangle $
and display the results arising from the five-point correlator  $\langle  \partial \ov Z  \sigma^+ \sigma^- \sigma^+ \sigma^- \rangle $ in the appendix \ref{app correlators one independent angle}.
From the limit $z \rightarrow 0$ one obtains
%
\begin{align}
\langle  \tau^+_{\alpha} (0) \sigma^-_{\alpha} (x) \sigma^+_{\alpha} (1) \sigma^-_{\alpha} (\infty) \rangle &=- x^{-\alpha(2-\alpha)}(1-x)^{-\alpha(1-\alpha)} \\& \hspace{-40mm}\times
{\frac{ \sin^\frac{3}{2}(\pi \alpha)}{2 \pi\,  _2F_1^\frac{1}{2}[\alpha,1-\alpha,1,x]\,    _2F^\frac{3}{2}_1[\alpha,1-\alpha,1,1-x]} } 
\sum_{p,q} \left( {v_b+e^{i\pi \alpha} v_a  \tau(x)} \right) \,e^{-S_{cl}} \,\,,\nn
\end{align}
which after reinstating all $x_i$ dependence gives (recall $x=\frac{x_{12} x_{34}}{x_{13} x_{24}}$)
\footnote{Here we use the fact that an open string four-point correlator is given by
 \begin{align}
{\cal A}_{open} = \langle \phi_{h_1}(x_1) \phi_{h_2}(x_2) \phi_{h_3}(x_3) \phi_{h_4}(x_4)\rangle ={\cal F}(x) \prod_{i<j} x_{ij}^{-(h_i  +h_j)+{\Delta  \over 3}}\,\,,
\end{align}
where $\Delta = \sum^4_{i=1} h_i$ and $x=\frac{x_{12} x_{34} }{x_{13} x_{24}}$.  This expression allows us to reinstate the $x_i$ dependence in all the four-point correlators. }
%
%
\begin{align}
\langle  \tau^+_{\alpha} (x_1) \sigma^-_{\alpha} (x_2) \sigma^+_{\alpha} (x_3) \sigma^-_{\alpha} (x_4) \rangle &= -
\left( \frac{x_{24}}{x_{12} x_{14}} \right)^{\alpha (2-\alpha)}  \, \left( \frac{x_{13}}{x_{23} x_{34}} \right)^{\alpha(1-\alpha)}
\label{eq correaltor tau+} 
 \\ & \hspace{-40mm}\times
{\frac{ \sin^\frac{3}{2}(\pi \alpha)}{2 \pi\,  _2F_1^\frac{1}{2}[\alpha,1-\alpha,1,x]\,    _2F^\frac{3}{2}_1[\alpha,1-\alpha,1,1-x]} } 
\sum_{p,q} \left( {v_b+e^{i\pi \alpha} v_a  \tau(x)} \right)\,e^{-S_{cl}} \,\,.
\nn
\end{align}
%
%
Analogously we get for the limit $z \rightarrow x$ 
\begin{align}
\langle  \sigma^+_{\alpha} (x_1) \tau^-_{\alpha} (x_2) \sigma^+_{\alpha} (x_3) \sigma^-_{\alpha} (x_4) \rangle &= -
\left( \frac{x_{13}}{x_{12} x_{23}} \right)^{(1-\alpha) (1+\alpha)}  \, \left( \frac{x_{24}}{x_{14} x_{34}} \right)^{\alpha(1-\alpha)}
\label{eq correaltor tau-}
 \\ & \hspace{-40mm}\times
{\frac{ \sin^\frac{3}{2}(\pi \alpha)}{2 \pi\,  _2F_1^\frac{1}{2}[\alpha,1-\alpha,1,x]\,    _2F^\frac{3}{2}_1[\alpha,1-\alpha,1,1-x]} } 
\sum_{p,q} \left( {v_b+e^{i\pi \alpha} v_a  \tau(x)} \right)\,e^{-S_{cl}} \,\,.
\nn
\end{align}
%
%
A comparison of \eqref{eq correaltor tau+}and \eqref{eq correaltor tau-} confirms the following identifications\footnote{In order to check that identification one takes \eqref{eq correaltor tau-} and exchanges $\alpha \rightarrow 1 - \alpha$ and simultaneously exchanges $x_1 \leftrightarrow x_2$ and $x_3 \leftrightarrow x_4$ and proves it gives the expression \eqref{eq correaltor tau+}. Note that the latter replacement, $x_1 \leftrightarrow x_2$ and $x_3 \leftrightarrow x_4$ , is in agreement with the non-vanishing of the trace over Chan-Paton factors accompanying each vertex operator, and thus each bosonic twist field correlator.} 
\begin{align}
\tau^-_{\alpha} = \tau^+_{1-\alpha} 
\end{align}
that were already suggested in \cite{Anastasopoulos:2011gn,Anastasopoulos:2011hj}  by studying the OPE's of such higher bosonic twist fields. Note that this identification uses the fact that $e^{i \pi \alpha} v_a =  e^{i \pi (1-\alpha)} v_a$ and holds also true for ${\widetilde \tau }^-_{\alpha} =  {\widetilde \tau }^+_{1-\alpha}$ and as we will see later also generalizes to higher excited bosonic twist fields. Let us point out that this identification among bosonic twist fields and its counterpart, the bosonic anti-twist fields, can be also checked by performing the same analysis with the bosonic twist field ordering of the type $\langle \sigma^-_{\alpha} (x_1) \sigma^+_{\alpha} (x_2) \sigma^-_{\alpha} (x_3) \sigma^+_{\alpha} (x_4) \rangle$ instead of the considered one here $\langle \sigma^+_{\alpha} (x_1) \sigma^-_{\alpha} (x_2) \sigma^+_{\alpha} (x_3) \sigma^-_{\alpha} (x_4) \rangle$.

The four point-correlator arising from the five-point correlator \eqref{eq dZ s s s s one angle} can be derived in an analogous fashion and is displayed in appendix \ref{app correlators one independent angle}.

\subsubsection{Four-point correlators containing two excited twist fields}

Analogously to the three-point correlators containing two excited bosonic twist fields the four-point correlators containing two excited bosonic twist fields arise from six-point correlators containing two conformal fields $\partial Z $ and $\partial \ov Z$. More specifically, they arise from the six-point correlators 
\begin{align} \nn
\langle \partial Z(z) \partial \ov Z(w)\sigma^+_{\a}(x_1) \sigma^-_{\a}(x_2) \sigma^+_{\a}(x_3) \sigma^-_{\a}(x_4)\rangle \\ \label{eq six point one angle}
\langle \partial  \ov Z(z) \partial  \ov Z(w)\sigma^+_{\a}(x_1) \sigma^-_{\a}(x_2) \sigma^+_{\a}(x_3) \sigma^-_{\a}(x_4) \rangle \\
\langle \partial   Z(z) \partial  Z(w)\sigma^+_{\a}(x_1) \sigma^-_{\a}(x_2) \sigma^+_{\a}(x_3) \sigma^-_{\a}(x_4) \rangle\,\,. \nn
\end{align} 
In contrast to the five-point correlators considered in the previous subsection \eqref{sec four-point one exc one angle} the pure quantum part of the correlators \ref{eq six point one angle} does not vanish. We will determine it applying the same procedure performed in section \ref{sec three-point two excited} for the derivation of the pure quantum five-point correlators.

Let us define the following three functions
\bea
g(z,w)&=& {\langle \partial Z_{qu}(z) \partial \ov Z_{qu}(w)\sigma^+_{\a}(x_1) \sigma^-_{\a}(x_2) \sigma^+_{\a}(x_3) \sigma^-_{\a}(x_4) \rangle 
\over \langle \sigma^+_{\a}(x_1) \sigma^-_{\a}(x_2) \sigma^+_{\a}(x_3) \sigma^-_{\a}(x_4)\rangle}\\
k(z,w) &=& {\langle \partial  \ov Z_{qu}(z) \partial  \ov Z_{qu}(w)\sigma^+_{\a}(x_1) \sigma^-_{\a}(x_2) \sigma^+_{\a}(x_3) \sigma^-_{\a}(x_4) \rangle 
\over \langle \sigma^+_{\a}(x_1) \sigma^-_{\a}(x_2) \sigma^+_{\a}(x_3) \sigma^-_{\a}(x_4)\rangle}\\
m(z,w) &=& {\langle \partial   Z_{qu}(z) \partial  Z_{qu}(w)\sigma^+_{\a}(x_1) \sigma^-_{\a}(x_2) \sigma^+_{\a}(x_3) \sigma^-_{\a}(x_4) \rangle 
\over \langle \sigma^+_{\a}(x_1) \sigma^-_{\a}(x_2) \sigma^+_{\a}(x_3) \sigma^-_{\a}(x_4) \rangle}
\eea
that will be determined momentarily using local behaviour of the bosonic twist fields as well as global monodromies. Together with the knowledge of the four-point correlator $\langle \sigma^+_{\a}(x_1) \, \sigma^-_{\a}(x_2)\, \sigma^+_{\a}(x_3)\, \sigma^-_{\a}(x_4) \rangle$ the functions $g(z,w)$, $k(z,w)$ and $m(z,w)$ allow us to derive the requested six-point correlators.  

Using the local behaviour of the bosonic twist fields \eqref{eq OPE twist fields} as well as \eqref{eq OPE partial Z} we obtain for the functions $g(z,w)$, $k(z,w)$ and $m(z,w)$
\begin{align}
g(z,w) &=\omega_{1-\a,1-\a}(z) \omega_{\a,\a}(w) \left\{{P\over (z-w)^2} +A\left(\{x_i \}\right)\right\}
\label{eq ansatz g}\\
k(z,w) &= \omega_{\a,\a}(z) \omega_{\a,\a}(w) B\left(\{x_i \}\right)
\label{eq ansatz k}\\
m(z,w) &= \omega_{1-\a,1-\a}(z) \omega_{1-\a,1-\a}(w) C\left(\{x_i \}\right)
\label{eq ansatz m}
\end{align}
where we use various symmetries of the functions under the exchanges of the $x_i$'s. Here $\omega_{\a,\b}$ is given by eq. \eqref{eq omega definition} and $P$ takes the form
\bea
2 P &=& (1-\xi) ~ (z-x_1)(z-x_2)(w-x_3)(w-x_4)\nn\\
&&+ 2 (-\a+\xi)           ~ (z-x_1)(w-x_2)(z-x_3)(w-x_4)\nn\\
&&+   (1-\xi) ~ (z-x_1)(w-x_2)(w-x_3)(z-x_4)\nn\\
&&+   (1-\xi) ~ (w-x_1)(z-x_2)(z-x_3)(w-x_4)\nn\\
&&+  2 (-1+\a+\xi)        ~ (w-x_1)(z-x_2)(w-x_3)(z-x_4)\nn\\
&&+   (1-\xi) ~ (w-x_1)(w-x_2)(z-x_3)(z-x_4)\,\,.
\eea
The $\xi$ is a free parameter that is neither fixed by the local behavior nor by any symmetry of the functions. The functions $A\left(\{x_i\}\right)$, $B\left(\{x_i\}\right)$ and $C\left(\{x_i\}\right)$ are functions on positions of the bosonic twist fields only and thus independent of $z$ and $w$. Those functions will be determined below using constraints arising from global monodromies.

From the boundary conditions \eqref{eq boundary conditions} one gets the monodromy constraints that result into the following set of equations\footnote{The setup with four bosonic twist field insertions has two independent world-sheet contours which we choose without loss of generality to be $0$ to $x$ and $x$ to $1$.}
\begin{align}
&\int^1_{x} dz \, \big(  g(z,w) -  k(z,w)  \big)=0  \label{eq monodromies 1}
~,~~  \int^x_{0} dz \, \big( e^{i \pi \a} g(z,w) - e^{-i \pi \a}   k(z,w)  \big)=0\\
&\int^1_{x} dw \big( m(z,w) - g (z,w)\big) =0
~,~~  \int^x_{0} dw\big( e^{i \pi \a}  m(z,w) - e^{-i\pi \a} g(z,w) \big)=0\,
\label{eq monodromies 2}
\end{align}
Focusing on the derivation of $g(z,w)$ and $k(z,w)$ the constraints \eqref{eq monodromies 1} lead after 
using  $SL(2,\mathbf{R})$ symmetry to fix $x_1 =0 $, $x_2=x$, $x_3=1$ and $x_4= x_{\infty}$, dividing by
$\omega_{\a,\a}(w)$ and taking the limit $w \rightarrow \infty$ to
\begin{align}
\int^1_{x} dz \, \big(  {\widetilde g}(z) - {\widetilde k} (z)  \big)=0 ~~,~~~~ \int^x_{0} dz \, \big( e^{i \pi \a} {\widetilde g}(z) - e^{-i \pi \a} {\widetilde k} (z)  \big) =0
\label{eq monodromy g and k}
\end{align}
where $\widetilde{g}(z)$ and $\widetilde{k}(z)$ are given by
\begin{align*} 
\widetilde{g}(z) &=x^{1+\a}_{\infty}\, \widetilde{\omega}_{1-\a,1-\a}(z) \left\{\frac{ (1-\xi) }{2}z +\frac{ (1-\xi) }{2}(z-1) -(1-\a-\xi) (z-x) + \widetilde{A}(x)\right\}\\
 \widetilde k(z) &=
 x^{1+\a}_{\infty} \widetilde{B}(x) \widetilde{\omega}_{\a,\a}(z)
\end{align*}
with
\begin{align}
 \widetilde{A}(x) =\frac{- A(0,x,1,x_{\infty})}{x_{\infty}} \qquad   \widetilde{B}(x) =\frac{- B(0,x,1,x_{\infty})}{x_{\infty}}\qquad 
\widetilde{w}_{\a,\b} (z) = z^{-\a} (z-x)^{\a-1} (z-1)^{-\b}
\label{eq definition of widetilde}
\end{align}
Solving the system of equations \eqref{eq monodromy g and k} by using the integrals displayed in appendix \ref{app integrals} we obtain for $ \widetilde{A}(x)$
\begin{align}
\widetilde{A} (x)=\frac{\a+\xi -1}{2} (2x-1) - \frac{\a}{2} \frac{_2F_1[-\a,\a,1,1-x]}{_2F_1[1-\a,\a,1,1-x]} +  \frac{\a}{2} \frac{_2F_1[-\a,\a,1,x]}{_2F_1[1-\a,\a,1,x] } 
\label{eq A via g and k}
\end{align} 
and   for $ \widetilde{B}(x)$
\begin{align}
\widetilde{B}(x) = \frac{\a}{2} e^{2i \pi \a} \left\{ 1-  \frac{_2F_1[-\a,\a,1,1-x]}{_2F_1[1-\a,\a,1,1-x]} -  \frac{_2F_1[-\a,\a,1,x]}{_2F_1[1-\a,\a,1,x]}  \right\}\,\,.
\end{align}
Together with \eqref{eq ansatz g} and \eqref{eq ansatz k} this determines the functions $g(z,w)$ and $k(z,w)$ at the positions $(z,w,0,x,1,\infty)$. Note that $g(z,w)$ and $k(z,w)$ are completely independent on the choice of $\xi$, as it should be. Thus for the function $g(z,w)$, $\widetilde{A}$ compensates for the $\xi$ dependence in $P$.

In an analogous way one can determine $\widetilde{C}(x)$ \footnote{
\label{footnote Ctilde}Here $\widetilde{C}(x) =\frac{- C(0,x,1,x_{\infty})}{x_{\infty}}  $ } by solving the system of equations \eqref{eq monodromies 2} and obtain for $\widetilde{A}(x)$ again \eqref{eq A via g and k} while for $\widetilde{C}(x)$ one gets
\bea
\widetilde{C}(z) =\frac{1-\a}{2} e^{-2 \pi i \a} \left\{1 -  \frac{ _2F_1[1-\a,\a-1,1,1-x] }{ _2F_1[1-\a,\a,1,1-x]} -
\frac{ _2F_1[1-\a,\a-1,1,x] }{ _2F_1[1-\a,\a,1,x]}
\right\} ~~~~~
\eea

The knowledge of the four-point correlator \eqref{eq 4 sigmas 1 angle}  then allows us to determine the pure quantum part of the six-point correlators $\langle \partial   Z  \partial \ov  Z \sigma^+_{\a} \sigma^-_{\a} \sigma^+_{\a} \sigma^-_{\a} \rangle$,  $\langle \partial \ov    Z  \partial \ov  Z \sigma^+_{\a} \sigma^-_{\a} \sigma^+_{\a}\sigma^-_{\a} \rangle$ and $\langle \partial    Z \partial   Z \sigma^+_{\a} \sigma^-_{\a} \sigma^+_{\a}\sigma^-_{\a} \rangle$. After a few manipulations one obtains 
%
%
 \begin{align}
& \langle \partial Z_{qu}(z) \partial \ov Z_{qu}(w)  \sigma^+_{\a} (0) \sigma^-_{\a}(x) \, \sigma^+_{\a} (1)\sigma^-_{\a} (\infty)\rangle \\ \nn
&~~~~ =  \frac{ \sqrt{\pi}  \left[x (1-x)\right]^{-\a(1-\a)} \left[ z (w-x) (z-1)\right]^{\a-1} \left[w (z-x) (w-1) \right]^{-\a}}{ \sqrt{\Gamma(\a) \Gamma(1-\a)\,   _2F_1[1-\a,\a,1,x] \, _2F_1[1-\a,\a,1,1-x]}} \\
& \hspace{20mm}\times\left\{ (1-\a) \frac{ z \,( w-x) \, (z-1)}{(z-w)^2} + \a\frac{ w \, (z-x) \, (w-1) }{(z-w)^2}
 \right. \nn \\
&  \left. \hspace{23mm}
+\frac{1}{2}  \a (1-\a) \left(x \frac{_2F_1[1-\a,\a,2,x]}{_2F_1[1-\a,\a,1,x] }  - (1-x) \frac{_2F_1[1-\a,\a,2,1-x]}{_2F_1[1-\a,\a,1,1-x]}  \right)
\right\} \,\,. \nn
 \end{align}
 For $\langle  \partial  \ov Z_{qu}  \partial \ov  Z_{qu} \sigma^+_{\a} \sigma^-_{\a} \sigma^+_{\a} \sigma^-_{\a} \rangle$ one obtains
%
%
\begin{align}
& \langle \partial \ov Z_{qu}(z) \partial \ov Z_{qu}(w)  \sigma^+_{\a} (0) \sigma^-_{\a}(x) \, \sigma^+_{\a} (1)\sigma^-_{\a} (\infty)\rangle \\ \nn 
& ~~~~~~~~~~~ = \a ~ e^{2i \pi \a} \frac{ \sqrt{\pi} \left[x (1-x)\right]^{-\a(1-\a)} \left[z (z-1) w (w-1) \right]^{-\a} \left[ (z-x)(w-x)\right]^{\a-1}
}{ 2 \sqrt{\Gamma(\a) \Gamma(1-\a)\, _2F_1[1-\a,\a,1,x] \, _2F_1[1-\a,\a,1,1-x]}} 
\\& \hspace{30mm}\times \left\{ 1-  \frac{_2F_1[-\a,\a,1,1-x]}{_2F_1[1-\a,\a,1,1-x]} -  \frac{_2F_1[-\a,\a,1,x]}{_2F_1[1-\a,\a,1,x]}  \right\} \nn
\end{align}
while $\langle \partial    Z_{qu}  \partial   Z_{qu} \sigma^+_{\a} \sigma^-_{\a} \sigma^+_{\a}\sigma^-_{\a} \rangle$ reads
%
%
\begin{align}
&\langle \partial Z_{qu}(z) \partial Z_{qu}(w)  \sigma^+_{\a} (0) \sigma^-_{\a}(x) \, \sigma^+_{\a} (1)\sigma^-_{\a} (\infty)\rangle\\ \nn
& ~~~~~ =\left(1-\a\right) e^{-2 \pi i \a} \frac{ \sqrt{\pi} \left[x(1-x)\right]^{-\a(1-\a)} \left[ z (z-1) \, w \, (w-1) \right]^{-\a} \left[ (z-x) (w-x) \right]^{\a-1}
}{ 2 \sqrt{\Gamma(\a) \Gamma(1-\a)\, _2F_1[1-\a,\a,1,x] \, _2F_1[1-\a,\a,1,1-x]}} \\ 
& \hspace{29mm}\times \left\{1 -  \frac{ _2F_1[1-\a,\a-1,1,1-x] }{ _2F_1[1-\a,\a,1,1-x]} -
\frac{ _2F_1[1-\a,\a-1,1,x] }{ _2F_1[1-\a,\a,1,x]}
\right\}\,\,.  \nn
\end{align}
Adding the ``\emph{mixed}" part that contains the classical solutions $\partial Z_{cl}$ and $\partial \ov Z_{cl}$, given by \eqref{eq classical sol one independent} as well as the world-sheet instanton contributions \eqref{eq WS torus one angle} we obtain for the three different six-point correlators:
%
%
 \begin{align}
  \label{eq six point g}
& \langle \partial Z(z) \partial \ov Z(w)  \sigma^+_{\a} (0) \sigma^-_{\a}(x) \, \sigma^+_{\a} (1)\sigma^-_{\a} (\infty)\rangle \\ \nn
&~~~~ =  \frac{  \sqrt{\pi}  \left[x (1-x)\right]^{-\a(1-\a)} \left[ z (w-x) (z-1)\right]^{\a-1} \left[w (z-x) (w-1) \right]^{-\a}}{ \sqrt{\Gamma(\a) \Gamma(1-\a)\, _2F_1[1-\a,\a,1,x] \, _2F_1[1-\a,\a,1,1-x]}} \\ 
& \hspace{6mm}\times \sum_{p,q} \left\{ (1-\a) \frac{  z \,( w-x) \, (z-1)}{(z-w)^2} + \a\frac{ w \, (z-x) \, (w-1) }{(z-w)^2} -\frac{\sin^2(\pi \a)}{ 4 \pi^2}\frac{v^2_b - e^{2 i \pi \a} v^2_a \tau^2(x)}{_2F_1[1-\a,\a,1,1-x]^2}
 \right.  \nn\\
&  \left. \hspace{23mm}
+\frac{1}{2}  \a (1-\a) \left(x \frac{_2F_1[1-\a,\a,2,x]}{_2F_1[1-\a,\a,1,x] }  - (1-x) \frac{_2F_1[1-\a,\a,2,1-x]}{_2F_1[1-\a,\a,1,1-x]}  \right)  \nn
\right\}  e^{-S_{cl}} \nn
 \end{align}
 \begin{align} 
 \label{eq six point k}
& \langle \partial \ov Z(z) \partial \ov Z(w)  \sigma^+_{\a} (0) \sigma^-_{\a}(x) \, \sigma^+_{\a} (1)\sigma^-_{\a} (\infty)\rangle \\ \nn 
& ~~~~~~~~~~~ = \a ~ e^{2i \pi \a} \frac{ \sqrt{\pi}  \left[x (1-x)\right]^{-\a(1-\a)} \left[z (z-1) w (w-1) \right]^{-\a} \left[ (z-x)(w-x)\right]^{\a-1}
}{ 2\sqrt{\Gamma(\a) \Gamma(1-\a)\, _2F_1[1-\a,\a,1,x] \, _2F_1[1-\a,\a,1,1-x]}} 
\\& \hspace{20mm}\times  \sum_{p,q} \left\{ 1-  \frac{_2F_1[-\a,\a,1,1-x]}{_2F_1[1-\a,\a,1,1-x]} -  \frac{_2F_1[-\a,\a,1,x]}{_2F_1[1-\a,\a,1,x]} 
\right.   \nn \\
& \hspace{70mm} \left.
 +\frac{ \sin^2(\pi \a)} {2 \pi^2 \a} \frac{\left( v_b - e^{i\pi \a} v_a \tau(x)\right)^2  } {_2F_1[\a,1-\a,1,1-x]^{2} }   
 \right\}  e^{-S_{cl}} \nn
\end{align}
\begin{align} 
 \label{eq six point m}
&\langle \partial Z(z) \partial Z(w)  \sigma^+_{\a} (0) \sigma^-_{\a}(x) \, \sigma^+_{\a} (1)\sigma^-_{\a} (\infty)\rangle\\ \nn
& ~~~~~~~~~~~ =\left(1-\a\right) e^{-2 \pi i \a} \frac{  \sqrt{\pi}  \left[x(1-x)\right]^{-\a(1-\a)} \left[ z (z-1) \, w \, (w-1) \right]^{\a-1} \left[ (z-x) (w-x) \right]^{-\a}
}{ 2\sqrt{\Gamma(\a) \Gamma(1-\a)  \, _2F_1[1-\a,\a,1,x] \, _2F_1[1-\a,\a,1,1-x]}} \\   \nn
& \hspace{29mm}\times  \sum_{p,q}  \left\{1 -  \frac{ _2F_1[1-\a,\a-1,1,1-x] }{ _2F_1[1-\a,\a,1,1-x]} -
\frac{ _2F_1[1-\a,\a-1,1,x] }{ _2F_1[1-\a,\a,1,x]} 
\right.  \nn \\  \nn
& \hspace{70mm} \left.
 +\frac{ \sin^2(\pi \a)} {2 \pi^2 (1-\a)} \frac{\left( v_b + e^{i\pi \a} v_a \tau(x)\right)^2  } {_2F_1[\a,1-\a,1,1-x]^{2} }   
 \right\}  e^{-S_{cl}} 
\end{align}
with $S_{cl}$ being the world-sheet instanton contribution given in equation \eqref{eq WS torus one angle}.

Given these six-point correlators we can derive the various four-point correlators containing higher excited twist fields by taking the appropriate limits as we will examplify momentarily. Taking the limit $z \rightarrow 0$ and $w \rightarrow x$ of \eqref{eq six point g} and
using the OPE's \eqref{eq OPE twist fields} one gets
\begin{align}
 \label{eq correlator +}
&\langle  \tau^+_{\a} (0)  \widetilde{\tau}^-_{\a}(x) \, \sigma^+_{\a} (1)\sigma^-_{\a} (\infty)\rangle \\
&~~~~ =  \frac{\sqrt{\pi}~  x^{-\a(3-\a)} (1-x)^{-\a(2-\a)}}{\sqrt{\Gamma(\a)\, \Gamma(1-\a) \, _2F_1[\a,1-\a,1,x]\, _2F_1[\a,1-\a,1,1-x]}}\nn\\ 
&~~~~~~ \times  \sum_{p,q}   \left\{\a(1-x) +\frac{1}{2} \a(1-\a) \left( x \frac{_2F_1[1-\a,\a,2,x]}{_2F_1[1-\a,\a,1,x]} - (1-x) \frac{_2F_1[1-\a,\a,2,1-x]}{_2F_1[1-\a,\a,1,1-x]} \right)  \nn \right. \\
&  \hspace{70mm} \left. 
-\frac{\sin^2(\pi \a)}{ 4 \pi^2}\frac{v^2_b - e^{2 i \pi \a} v^2_a \tau^2(x)}{_2F_1[1-\a,\a,1,1-x]^2}
 \right\}\nn e^{-S_{cl}} 
\end{align}
which after reinstating all $x_i$ dependence gives (recall $x=\frac{x_{12} x_{34}}{x_{13} x_{24}}$)
\footnote{Here we use the fact that an open string four-point correlator is given by
 \begin{align}
{\cal A}_{open} = \langle \phi_{h_1}(x_1) \phi_{h_2}(x_2) \phi_{h_3}(x_3) \phi_{h_4}(x_4)\rangle ={\cal F}(x) \prod_{i<j} x_{ij}^{-(h_i  +h_j)+{\Delta  \over 3}}
\end{align}
where $\Delta = \sum^4_{i=1} h_i$ and $x=\frac{x_{12} x_{34} }{x_{13} x_{24}}$.  This expression allows us to reinstate the $x_i$ dependence in all the four-point correlators. }
\begin{align} 
&\langle  \tau^+_{\a} (x_1)  \widetilde{\tau}^-_{\a}(x_2) \,  \sigma^+_{\a} (1)\sigma^-_{\a} (\infty)\rangle  \\ &\hspace{5mm} =
\frac{\sqrt{\pi}
~
x^{-\a(3-\a)}_{12} \, x^{-\a(1-\a)}_{34} \,\left(\frac{x_{13} \, x_{24}}{ x_{14}\,x_{23}}\right)^{\a(2-\a)}
}{\sqrt{\Gamma(\a)\, \Gamma(1-\a)\,_2F_1[\a,1-\a,1,x]\, _2F_1[\a,1-\a,1,1-x]}}
\nn \\ &\hspace{10mm}  \times  \sum_{p,q}  \left\{\a (1-x) +\frac{1}{2} \a (1-\a) \left( x \frac{_2F_1[1-\a,\a,2,x]}{_2F_1[1-\a,\a,1,x]} - (1-x) \frac{_2F_1[1-\a,\a,2,1-x]}{_2F_1[1-\a,\a,1,1-x]} \right) 
 \nn \right. \\
&  \hspace{70mm} \left. 
-\frac{\sin^2(\pi \a)}{ 4 \pi^2}\frac{v^2_b - e^{2 i \pi \a} v^2_a \tau^2(x)}{_2F_1[1-\a,\a,1,1-x]^2}
 \right\}\nn  e^{-S_{cl}}  \,\,.
\end{align}
If we take the limit  $z\rightarrow x$ and $w \rightarrow 0$ of \eqref{eq six point g} we get the four-point correlator $\langle  \widetilde{\tau}^+_{\a}  \tau^-_{\a}\,  \sigma^+_{\a} \sigma^-_{\a}  \rangle$. Comparing this correlator to \eqref{eq correlator +} confirms again the identifications     
\begin{align}
\tau^-_{\a} = \tau^+_{1-\a} \qquad \qquad  {\widetilde \tau }^-_\a =  {\widetilde \tau }^+_{1-\a}
\label{eq identifications}
\end{align}
that were already suggested in chapter \ref{sec 4-point one angle}. Looking at other limits of the six-point correlators  \eqref{eq six point g}, \eqref{eq six point k} and \eqref{eq six point m} that give us 
four-point correlators containing excited bosonic twist fields we can generalize this  
identification to the higher excited twist fields $\omega^-$ and $\widetilde \omega^-$. Thus we have
\begin{align}
\omega^-_{\a} = \omega^+_{1-\a} \qquad \qquad  {\widetilde \omega }^-_\a =  {\widetilde \omega}^+_{1-\a}\,\,.
\end{align}
In appendix \ref{app correlators one independent angle} we display all other four-point correlators arising from the six-point correlators \eqref{eq six point g}, \eqref{eq six point k} and \eqref{eq six point m} by investigation of particular limits of $z$ and  $w$.

\subsection{Four-point correlators with two independent angles}
In this section we perform the same analysis as above, however for the more generalized setup of two independent angles. The bosonic twist field insertions $\sigma^+_{\a}(x_1)$, $\sigma^-_{\a}(x_2)$, $\sigma^+_{\b}(x_3)$ and $\sigma^-_{\b}(x_4)$ lead to the following boundary conditions  
\begin{align}
\partial Z -  \partial \ov Z = 0& \,\,\,\,\, \text{for} \,\,\,\,(-\infty, x_1) \cup (x_2,x_3) \cup  (x_4, \infty)\nn\\
e^{i \pi \alpha}\partial Z - e^{-i\pi \alpha} \partial \ov Z = 0&  \,\,\,\,\, \text{for} \,\,\,\,(x_1,x_2)
\,\,\nn\\e^{i \pi \b}
\partial Z - e^{-i\pi \b} \partial \ov Z = 0&  \,\,\,\,\, \text{for} \,\,\,\,(x_3,x_4) \,\,.
\end{align}
Analogously to section \ref{sec 4-point one angle} we will determine the correlators containing
one or two excited bosonic twist fields via deriving five- and six-point correlators. Let us start with the derivation of the four-point correlators containing one excited bosonic twist field.

\subsubsection{Four-point correlators containing one excited twist field}
As in the subsection \ref{sec four-point one exc one angle} the four-point correlators containing one excited bosonic twist field arise from the five-point correlators  
\begin{align}
\langle \partial Z(z)  \sigma^+_{\alpha} (x_1) \sigma^-_{\alpha} (x_2) \sigma^+_{\b} (x_3) \sigma^-_{\b} (x_4) \rangle  ~~\\
\langle \partial  \ov Z(z) \sigma^+_{\alpha} (x_1) \sigma^-_{\alpha} (x_2) \sigma^+_{\b} (x_3) \sigma^-_{\b} (x_4)  \rangle \,\,,
\label{eq 5 point dZ correlator ab}
\end{align} 
which has no purely quantum part, but is completely dictated by the classical part of $\partial Z$ and $\partial  \ov Z$. They have been computed in \cite{Abel:2003yx} (see also \cite{Cvetic:2003ch}) and take the form
\begin{align} 
\partial Z_{cl} (z) = 
 \widetilde{a} ~\omega_{1-\alpha,1-\b}(z) \qquad \qquad \partial \ov Z_{cl} (z) =
 ~ \widetilde{b} ~\omega_{\alpha,\b}(z) \,\,.
\label{eq classical sol two independent ab}
\end{align}
with $\omega_{\alpha,\beta}$ being \eqref{eq omega definition} and $\widetilde{a} $ and $\widetilde{b}$ given by
\bea
&&\widetilde{a} =- \Big(v_{b} G_2[x]+  \frac{\sin(\pi \a)}{\pi} e^{i \pi \a} v_{a} (1-x)^{\a-\b} B_1  H_1[1-x]\Big) I^{-1}(x) \nn\\
&&\widetilde{b} =- \Big(v_{b} G_1[x] -  \frac{\sin(\pi \a)}{\pi} e^{i \pi \a} v_{a} (1-x)^{\b-\a} B_2  H_2[1-x]\Big) I^{-1}(x)\,\,.
\eea
%
%
Here $v_a$ and $v_b$ are the same as in the previous subsection \ref{sec 4-point one angle}
and we used the following definitions
\begin{align}
& B_1=\frac{\Gamma(\a)\,\Gamma(1-\b)}{\Gamma(1+\a-\b)} && B_2=\frac{\Gamma(\b)\,\Gamma(1-\a)}{\Gamma(1+\b-\a)} \nn\\
& G_1[x]= {_2F}_1[\a,1-\b,1;x]     && G_2[x]= {_2F}_1[1-\a,\b,1;x] \label{eq def hypergeometric functions}\\
& H_1[x]= {_2F}_1[\a,1-\b,1+\a-\b;x] && H_2[x]={_2F}_1[1-\a,\b,1-\a+\b;x] \,\, \nn
\end{align}
with which $I(x)$ takes the form
\bea
I(x)= B_1\,G_2[x] H_1[1-x]+ B_2\, G_1[x] H_2[1-x]\,\,.
\label{eq def of I}
\eea
It is easy to check that the at the limit $\b\to \a$ the classical solution \eqref{eq classical sol two independent ab} simplifies to the one with just 
one independent angle \eqref{eq classical sol one independent}. Moreover in the limit $x \rightarrow 1$ this solution goes over into the classical solution of three bosonic twist field insertions \eqref{eq classical solution 3 point} discussed in chapter \ref{sec three-point}.

Together with the four point correlator \cite{Cvetic:2003ch, Abel:2003yx,Lust:2004cx,Lust:2008qc,Anastasopoulos:2011gn}
\begin{align}
\langle \sigma^+_{\alpha}(0) \, \sigma^-_{\alpha}(x)\, \sigma^+_{\b}(1)\, \sigma^-_{\b}(\infty) \rangle =
x^{-\a(1-\a)}(1-x)^{\a\b-\frac{\a}{2}-\frac{\b}{2}}\sqrt{\frac{2\pi}{I(x)}}
\label{eq 4 sigmas 2 angle} 
\end{align}
we obtain for the two five point correlators
\begin{align} \nn
&\langle\partial Z(z)   \sigma^+_\a (0) \sigma^-_{\a} (x) \sigma^+_\b (1) \sigma^-_{\b} (\infty) \rangle    = 
 - z^{-1+\a} (z-x)^{-\a}  (z-1)^{-1+\b}  ~
x^{-\a(1-\a)} (1-x)^{\a\b -\frac{\a}{2} -\frac{\b}{2}}   
\\
&   \hspace{8mm}  \times  \sqrt{ 2 \pi} e^{-\pi  i \b} 
 \sum_{p, q} \frac{\left( G_2[x] v_b +  \frac{\sin(\pi \a)}{\pi} e^{i \pi \a} v_a (1-x)^{\a-\b} B_1 H_1[1-x]  \right)}{I^{\frac{3}{2}}(x)} \,e^{-S_{cl}}
\label{eq dZ ssss ab} 
\end{align}
\begin{align}
&\langle   \partial \ov Z (z) \sigma^+_\a (0) \sigma^-_{\a} (x) \sigma^+_\b (1) \sigma^-_{\b} (\infty) \rangle =
 ~ z^{-\a} (z-x)^{-1+\a}  (z-1)^{-\b} ~
x^{-\a(1-\a)} (1-x)^{\a\b -\frac{\a}{2} -\frac{\b}{2}}    \nn
\\ 
& \hspace{8mm} \times \sqrt{ 2 \pi}    e^{\pi  i \b} 
 \sum_{p, q} 
\frac{ \left( G_1[x] v_b -  \frac{\sin(\pi \a)}{\pi} (1-x)^{\b-\a} B_2 H_2[1-x] e^{i \pi \a} v_a \right)}{I^{\frac{3}{2}}(x)} \,e^{-S_{cl}}\,\,.
\label{eq ov dZ ssss ab}
\end{align}
The classical action can be computed in an analogous fashion as previously using \eqref{eq classical action} as well as \eqref{eq classical sol two independent ab} which gives \cite{Abel:2003vv} (see also \cite{Lust:2008qc})
\begin{align}
S_{cl}^{T^2}= \frac{\pi}{\alpha'}
\sin(\pi \a) \left\{ |v_a|^2 \left(\tau(x) +\frac{\rho}{2}\right) 
+ |v_b+\frac{\rho}{2}v_a|^2 \left(\tau(x) +\frac{\rho}{2}\right)^{-1}
\right\}
\label{eq WS four 2 ind angles}
\end{align}
where $\tau(x)$ and $\rho$ are
\bea
\tau(x)=\frac{1}{\pi} \sin(\pi \a) \frac{B_2  H_2[1-x]}{G_2[x]}
~~,~~~~~~~\rho=-\frac{\sin\left(\pi (\a-\b)\right)}{\sin(\pi \b)}   
\eea
and  $v_a$ and $v_b$ are defined as previously. Again we have to sum over all world-sheet instanton contribution, i.e. sum over all possible closed polygons connecting the four intersection points, which implies a sum over all integers $p$ and $q$.

Given those five-point correlators \eqref{eq dZ ssss ab} and \eqref{eq ov dZ ssss ab} we can derive the desired four-point correlators. We illustrate this by applying this procedure to  the  five-point correlator \eqref{eq dZ ssss ab}. The limit $z \rightarrow 0$ gives
\begin{align}
&\langle  \tau^+_{\alpha} (0) \sigma^-_{\alpha} (x) \sigma^+_{\b} (1) \sigma^-_{\b} (\infty) \rangle 
 = 
-\sqrt{ 2 \pi} ~ e^{-i\pi \a}  ~
x^{-\a(2-\a)} (1-x)^{\a\b -\frac{\a}{2} -\frac{\b}{2}}  \nn
\\
&  \hspace{35mm}\times   \sum_{p, q}  \frac{\left( G_2[x] v_b +  \frac{\sin(\pi \a)}{\pi} (1-x)^{\a-\b} B_1 H_1[1-x] e^{i \pi \a} v_a \right)}{I^{\frac{3}{2}}(x)} 
\label{eq ov tsss ab}
\end{align}
which after reinstating the whole $x_i$ dependence takes the form
\begin{align}
&\langle  \tau^+_{\alpha} (x_1) \sigma^-_{\alpha} (x_2) \sigma^+_{\b} (x_3) \sigma^-_{\b} (x_4) \rangle
=
-x_{12}^{-\a(2-\a)} \, x_{34}^{-\b(1-\b)} \left(\frac{x_{23}}{x_{13}} \right)^{\a\b -\frac{\a}{2} -\frac{\b}{2}} 
\left(\frac{x_{14}}{x_{24}} \right)^{\a\b -\frac{3 \a}{2} -\frac{\b}{2}} 
\nn
\\
&  \hspace{8mm} \times \sqrt{ 2 \pi} ~ e^{-i\pi \a}  \sum_{p, q}   \frac{\left( G_2[x] v_b +  \frac{\sin(\pi \a)}{\pi} (1-x)^{\a-\b} B_1 H_1[1-x] e^{i \pi \a} v_a \right)}{I^{\frac{3}{2}}(x)} \, e^{-S_{cl}}\,
\label{eq ov tsss ab all x1x2x3x4}
\end{align}
where we used the definitions \eqref{eq def hypergeometric functions} as well as \eqref{eq def of I} and 
the world-sheet instanton contributions are given by \eqref{eq WS four 2 ind angles}. The other independent four-point correlator arising from the five-point correlator $\langle \partial \ov Z \sigma \sigma \sigma \sigma \rangle$ is displayed in appendix \ref{app correlators two independent angle}.

\subsubsection{Four-point correlators containing two excited twist fields}

Four-point correlators containing two excited bosonic twist fields arise from the six-point correlators
\begin{align}
\langle \partial Z(z) \partial  Z (w)  \sigma^+_\a (x_1) \sigma^-_{\a} (x_2) \sigma^+_\b (x_3) \sigma^-_{\b} (x_4) \rangle \\ \label{eq six-point correlator}
\langle \partial  \ov Z(z) \partial \ov  Z (w)  \sigma^+_\a (x_1) \sigma^-_{\a} (x_2) \sigma^+_\b (x_3) \sigma^-_{\b} (x_4) \rangle \\ 
\langle \partial Z(z) \partial \ov  Z (w)  \sigma^+_\a (x_1) \sigma^-_{\a} (x_2) \sigma^+_\b (x_3) \sigma^-_{\b} (x_4) \rangle 
\end{align}  
which we derive momentarily, applying the same procedure as for the five-point correlators in section \ref{sec three-point two excited}.
Let us start again with the derivation of the pure quantum part, for which we define
in a similar fashion to the case with one independent angle
\bea
g(z,w) &=& {\langle \partial Z_{qu}(z) \partial \ov Z_{qu}(w)\sigma^+_\a (x_1) \sigma^-_{\a} (x_2) \sigma^+_\b (x_3) \sigma^-_{\b} (x_4) \rangle
\over \langle \sigma^+_\a (x_1) \sigma^-_{\a} (x_2) \sigma^+_\b (x_3) \sigma^-_{\b} (x_4) \rangle}\\
k(z,w) &=& {\langle \partial  \ov Z_{qu}(z) \partial  \ov Z_{qu}(w)\sigma^+_\a (x_1) \sigma^-_{\a} (x_2) \sigma^+_\b (x_3) \sigma^-_{\b} (x_4) \rangle 
\over \langle \sigma^+_\a (x_1) \sigma^-_{\a} (x_2) \sigma^+_\b (x_3) \sigma^-_{\b} (x_4) \rangle}\\
m(z,w) &=& {\langle \partial   Z_{qu}(z) \partial  Z_{qu}(w)\sigma^+_\a (x_1) \sigma^-_{\a} (x_2) \sigma^+_\b (x_3) \sigma^-_{\b} (x_4) \rangle
\over \langle\sigma^+_\a (x_1) \sigma^-_{\a} (x_2) \sigma^+_\b (x_3) \sigma^-_{\b} (x_4) \rangle}\,\,.
\eea   

Applying the OPE's \eqref{eq OPE twist fields} the ansatz's for the functions read
\bea
g(z,w) &=& \omega_{1-\a,1-\b}(z) \omega_{\a,\b}(w) \left\{{P\over (z-w)^2} +A\left(\{x_i \}\right)\right\}\\
k(z,w) &=& \omega_{\a,\b}(z) \omega_{\a,\b}(w) B\left(\{x_i \}\right)\\
m(z,w) &=& \omega_{1-\a,1-\b}(z) \omega_{1-\a,1-\b}(w) C\left(\{x_i \}\right) \,\,,
\eea
where $\omega_{\a,\b}$ is given by \eqref{eq omega definition} and $P$ takes the form
\bea
2P &=& 
\left(1-4 \xi \right)  (z-x_1)(z-x_2)(w-x_3)(w-x_4)\nn\\
&&+ 
 (1-\a-\b+2\xi ) (z-x_1)(w-x_2)(z-x_3)(w-x_4)\nn\\
&&- 
 (\a-\b-2\xi) (z-x_1)(w-x_2)(w-x_3)(z-x_4)\nn\\
&&+
 (\a-\b+2\xi) (w-x_1)(z-x_2)(z-x_3)(w-x_4)\nn\\
&&+ 
  (\a+\b-1+2\xi) (w-x_1)(z-x_2)(w-x_3)(z-x_4)\nn\\
&&+ 
 (1-4 \xi) (w-x_1)(w-x_2)(z-x_3)(z-x_4)\,\,.
\eea
Here we again used various symmetries of the function $g(z,w)$ as well as the singular behavior \eqref{eq OPE partial Z}. The parameter $\xi$ is not fixed by any symmetry or the singular behavior, but as before in the one independent angle case we expect that $g(z,w)$ itself is independent of $\xi$. Thus we expect that the $\xi$ dependence of $P$ is compensated by $A\left(\{x_i \}\right)$.
The functions $A\left(\{x_i \}\right)$, $B\left(\{x_i \}\right)$ and  $C\left(\{x_i \}\right)$ will be determined momentarily using the monodromy constraints that read again \eqref{eq monodromies 1} and \eqref{eq monodromies 2}. As before we fix the vertex operator positions to $x_1 =0 $, $x_2=x$, $x_3=1$ and $x_4= x_{\infty}$ using $SL(2,\mathbf{R})$ symmetry, divide the monodromy constraint \eqref{eq monodromies 1} by $\omega_{\a,\b}(w)$ and the monodromy constraint \eqref{eq monodromies 2} by $\omega_{1-\a,1-\b}(z)$ and eventually take the limit $w \rightarrow \infty$ and $z \rightarrow \infty$, respectively. This will lead to the monodromy constraints
\bea
&&\int^1_{x} dz \, \big(  {\widetilde g}(z) - {\widetilde k} (z)  \big)=0 ~~, \qquad \int^x_{0} dz \, \big( e^{i \pi \a} {\widetilde g}(z) - e^{-i \pi \a} {\widetilde k} (z)  \big) =0
\label{eq monodromy g and k 2 angles}\\
&&\int^1_{x} dz \, \big(  {\widetilde g}'(z) - {\widetilde m} (z)  \big)=0 ~~, ~~~~ \int^x_{0} dz \, \big( e^{i \pi \a} {\widetilde g}'(z) - e^{-i \pi \a} {\widetilde m} (z)  \big) =0~~~~~~
\label{eq monodromy g and m 2 angles}
\eea
with 
\begin{align}
&\widetilde{g}(z) = \widetilde{\omega}_{1-\a,1-\b}(z) \left\{ \frac{4 \xi-1+ (1-\a-\b-2 \xi)x+ 2\b z}{2} 
+ {\widetilde A}(x)\right\}\\
&\widetilde{g}'(z) = \widetilde{\omega}_{\a,\b}(z) \left\{\frac{4 \xi-1+(\a+\b-1-2 \xi)x + 2(\b-1)z}{2}  
+ {\widetilde A}(x)\right\}\\
&\widetilde{k}(z) = {\widetilde B } (x)\widetilde{\omega}_{\a,\b}(z) \qquad  \qquad \qquad \qquad \widetilde{m}(z) = {\widetilde C} (x)\widetilde{\omega}_{1-\a,1-\b}(z) \,\,.
\end{align}
Here ${\widetilde A}$, ${\widetilde B}$, ${\widetilde C}$ and $\widetilde{\omega}$ are defined as in the previous subsection (see eq. \eqref{eq definition of widetilde} and footnote \ref{footnote Ctilde}).
Solving the system of equations \eqref{eq monodromy g and k 2 angles} and \eqref{eq monodromy g and m 2 angles} gives
\bea
\widetilde{A} & =& \frac{1}{2} (1-4 \xi) + \frac{1}{2} (\a+\b-1+2\xi) x  - \a \\
&& ~~~ - \b\,  \frac{B_2 G_1[x] }{I(x)}  \, _2F_1[-\a,\b,1-\a+\b,1-x] + \a\, \frac{ B_2 G_1[x]  H_2[1-x]\, }{I(x)}\nn
\\ && ~~~  - \a (1-x)^{\b-\a} \frac{ B_2   H_2[1-x]\, }{I(x)} \, _2F_1[-\a,\b,1,x]  + \a (1-x)^{\b-\a}\, \frac{_2F_1[-\a,\b,1,x] }{G_1[x]}  \nn\\
\widetilde{B} &=&\b e^{2 \pi i \b} (1-x)^{\b-\a} \\
&& \left.   \times
\left\{\frac{\a}{\b} \frac{B_2  G_1[x] H_2[1-x]}{I(x)} \right. 
-\frac{B_2 G_1[x]}{I(x)} \, _2F_1[-\a,\b,1-\a+\b,1-x] 
 \right.  
\nn\\ 
&& \left.  ~~~~~~~~~~~~~~~~~~~~~~~~~~~~~~~~~~ 
- \frac{\a}{\b} (1-x)^{\b-\a} \frac{B_2 H_2[1-x]}{I(x)} \, _2F_1[-\a,\b,1,x] \right\}  \nn\\
\widetilde{C} &=&  (1-\a) e^{-2 \pi i \b} (1-x)^{\a-\b} \frac{B_1}{I(x)} \\
&&\times \Big\{
G_2[x] H_1[1-x] 
-\frac{1-\b}{1-\a} x G_2[x] \, _2F_1[2-\b,\a,1+\a-\b,1-x] \nn \\
&& ~~~~~~~~~~~~~~~~~~~~~~~~~~~~~~~~~~~ 
-  (1-x)  H_1[1-x]\, _2F_1[2-\a,\b,1,x] \Big\} \,\,, \nn
\eea
where we used the definitions of \eqref{eq def hypergeometric functions} and \eqref{eq def of I}. Note that as required the functions $\widetilde{g}(z)$ and $\widetilde{g}'(z)$ are indeed independent of the choice of $\xi$.

Together with the expression for the four point correlator $\langle \sigma^+_{\a} \sigma^-_{\a} \sigma^+_{\b} \sigma^-_{\b} \rangle$ \eqref{eq 4 sigmas 2 angle}  one obtains for the quantum six-point correlators
\begin{align}
\label{eq six point g ab}
& \langle \partial Z_{qu}(z) \partial \ov Z_{qu}(w)  \sigma^+_{\a} (0) \sigma^-_{\a}(x) \, \sigma^+_{\b} (1)\sigma^-_{\b} (\infty)\rangle \\ \nn
& = 
\frac{\sqrt{2 \pi} [z (w-x) ]^{\a-1} [w (z-x)]^{-\a} (z-1)^{\b-1} (w-1)^{-\b}  x^{-\a(1-\a)} (1-x)^{\a\b -\frac{\a}{2} -\frac{\b}{2}}}{\sqrt{I(x)}}\\ \nn
& 
  \times \frac{1}{2} \Big\{ \frac{z(z-x)(w-1) +(1-\a-\b)z(w-x)(z-1) - (\a-\b) z (w-x) (w-1)}{(z-w)^2}\\ \nn
& \hspace{3mm} +\frac{(\a-\b) w (z-x)(z-1) + (\a+\b-1) w (z-x) (w-1) + w (w-x) (z-1) }{(z-w)^2}\\ \nn
& \hspace{3mm} + 1-2\a +(\a+\b-1)x   + 2\, \frac{ B_2 G_1[x]   }{I(x)}  \big( \a H_2[1-x]- \b\,\,   _2F_1[-\a,\b,1-\a+\b,1-x] \big)\\ \nn
&\hspace{3mm} - 2\a (1-x)^{\b-\a} \frac{ B_2  H_2[1-x]\, }{I(x)} \, _2F_1[-\a,\b,1,x] 
+ 2\a (1-x)^{\b-\a}\, \frac{_2F_1[-\a,\b,1,x] }{G_1[x]}
\Big\} \nn
\end{align}
\begin{align}\label{eq six point k ab}
& \langle \partial \ov Z_{qu}(z) \partial \ov Z_{qu}(w)  \sigma^+_{\a} (0) \sigma^-_{\a}(x) \, \sigma^+_{\b} (1)\sigma^-_{\b} (\infty)\rangle \\ \nn
&=
\frac{\sqrt{2 \pi } [z\, w ]^{-\a} [(z-x)(w-x)]^{\a-1} [(z-1)(w-1)]^{-\b} x^{-\a(1-\a)} (1-x)^{\a\b -\frac{\a}{2} -\frac{\b}{2}} }{\sqrt{I(x)}} \\
&\times e^{2 \pi i \b} (1-x)^{\b-\a} \frac{B_2}{I(x)}  \Big\{ {\a}~ {G_1[x] H_2[1-x]} - {\b}~ {G_1[x]} \, _2F_1[-\a,\b,1-\a+\b,1-x]  \nn\\
&\hspace{50mm}- {\a}~ (1-x)^{\b-\a} {H_2[1-x]} \, _2F_1[-\a,\b,1,x] \Big\}
\nn
\end{align}
\begin{align}\label{eq six point m ab}
& \langle \partial Z_{qu}(z) \partial Z_{qu}(w)  \sigma^+_{\a} (0) \sigma^-_{\a}(x) \, \sigma^+_{\b} (1)\sigma^-_{\b} (\infty)\rangle\\
&=
\frac{\sqrt{2 \pi } [z\, w ]^{\a-1} [(z-x)(w-x)]^{-\a} [(z-1)(w-1)]^{\b-1} x^{-\a(1-\a)} (1-x)^{\a\b -\frac{\a}{2} -\frac{\b}{2}} }{\sqrt{I(x)}}\nn \\
%
%
%
& \times(1-\a) e^{-2 \pi i \b} (1-x)^{\a-\b} \frac{B_1}{I(x)} 
\Big\{
  -(1-x)  H_1[1-x]\, _2F_1[2-\a,\b,1,x] \nn\\
&\hspace{34mm} +G_2[x] H_1[1-x]    -\frac{1-\b}{1-\a} x G_2[x] \, _2F_1[2-\b,\a,1+\a-b,1-x]\Big\} \nn\,\,.
\end{align}
Adding the mixed part containing the classical solutions $\partial Z_{cl}$ and $\partial \ov Z_{cl}$ \eqref{eq classical sol two independent ab}
 as well as the world-sheet instantons \eqref{eq WS four 2 ind angles}
\begin{align}
\label{eq 6 point correlator Z  ov Z}
& \langle \partial Z(z) \partial \ov Z(w)  \sigma^+_{\a} (0) \sigma^-_{\a}(x) \, \sigma^+_{\b} (1)\sigma^-_{\b} (\infty)\rangle 
\\ \nn
& = 
\frac{\sqrt{2 \pi} [z (w-x) ]^{\a-1} [w (z-x)]^{-\a} (z-1)^{\b-1} (w-1)^{-\b}  x^{-\a(1-\a)} (1-x)^{\a\b -\frac{\a}{2} -\frac{\b}{2}}}{\sqrt{I(x)}}\\ \nn
& 
  \times \frac{1}{2} \sum_{p,q} \Big\{ \frac{z(z-x)(w-1) +(1-\a-\b)z(w-x)(z-1) - (\a-\b) z (w-x) (w-1)}{(z-w)^2}\\ \nn
& \hspace{3mm} +\frac{(\a-\b) w (z-x)(z-1) + (\a+\b-1) w (z-x) (w-1) + w (w-x) (z-1) }{(z-w)^2}\\ \nn
& \hspace{3mm} + 1-2\a +(\a+\b-1)x   + 2\, \frac{ B_2 G_1[x]   }{I(x)}  \big( \a H_2[1-x]- \b\,\,   _2F_1[-\a,\b,1-\a+b,1-x] \big)\\ \nn
&\hspace{3mm} - 2\a (1-x)^{\b-\a} \frac{ B_2  H_2[1-x]\, }{I(x)} \, _2F_1[-\a,\b,1,x] 
+ 2\a (1-x)^{\b-\a}\, \frac{_2F_1[-\a,\b,1,x] }{G_1[x]}\\
&  \hspace{3mm}  -\frac{\left( G_1[x] v_b +  \frac{\sin(\pi \a)}{\pi} B_1 H_1[1-x] e^{i \pi \a} v_a \right) \left( G_2[x] v_b -  \frac{\sin(\pi \a)}{\pi} B_2 H_2[1-x] e^{i \pi \a} v_a \right)}{I^2(x)} \nn
\Big\} e^{-S_{cl}} \nn
\end{align}
\begin{align}
\label{eq 6 point correlator ov Z ov Z}
& \langle \partial  \ov Z(z) \partial  \ov Z(w)  \sigma^+_{\a} (0) \sigma^-_{\a}(x) \, \sigma^+_{\b} (1)\sigma^-_{\b} (\infty)\rangle \\ \nn
&=
\frac{\sqrt{2 \pi } [z\, w ]^{-\a} [(z-x)(w-x)]^{\a-1} [(z-1)(w-1)]^{-\b} x^{-\a(1-\a)} (1-x)^{\a\b -\frac{\a}{2} -\frac{\b}{2}} }{\sqrt{I(x)}} \\
&\times e^{2 \pi i \b}  \sum_{p,q} \bigg\{   (1-x)^{\b-\a} \frac{B_2}{I(x)}  \Big\{ {\a}~ {G_1[x] H_2[1-x]} - {\b}~ {G_1[x]} \, _2F_1[-\a,\b,1-\a+\b,1-x] \nn\\
& 
\hspace{20mm}- {\a}~ (1-x)^{\b-\a} {H_2[1-x]} \, _2F_1[-\a,\b,1,x]\Big\} \nn \\ &  \hspace{20mm}+ 
\frac{\left( G_1[x] v_b -  \frac{\sin(\pi \a)}{\pi} (1-x)^{\b-\a} B_2 H_2[1-x] e^{i \pi \a} v_a \right)^2}{I^2(x)} 
\bigg\} e^{-S_{cl}}
\nn
\end{align}
\begin{align}
\label{eq 6 point correlator Z Z}
& \langle \partial  Z(z) \partial  Z(w)  \sigma^+_{\a} (0) \sigma^-_{\a}(x) \, \sigma^+_{\b} (1)\sigma^-_{\b}  (\infty)\rangle\\
&= \nn
\frac{\sqrt{2 \pi } [z\, w ]^{\a-1} [(z-x)(w-x)]^{-\a} [(z-1)(w-1)]^{\b-1} x^{-\a(1-\a)} (1-x)^{\a\b -\frac{\a}{2} -\frac{\b}{2}} }{\sqrt{I(x)}}\\
& \nn  \times e^{-2 \pi i \b}  \sum_{p,q}  \bigg\{  (1-\a)  (1-x)^{\a-\b} \frac{ B_1}{I(x)} 
\Big\{
 -(1-x)  H_1[1-x]\, _2F_1[2-\a,\b,1,x]  
\nn\\ & \nn
 \hspace{15mm} 
+G_2[x] H_1[1-x]    -\frac{1-\b}{1-\a} x G_2[x] \, _2F_1[2-\b,\a,1+\a-\b,1-x] \Big\} \\ &  \hspace{15mm}+ 
 \frac{ \left( G_2[x] v_b +  \frac{\sin(\pi \a)}{\pi} (1-x)^{\a-\b} B_1 H_1[1-x] e^{i \pi \a} v_a \right)^2}{I^2(x)} 
\bigg\} e^{-S_{cl}} \nn\,\,.
\end{align}
Note that taking the limits $x\to 1$ of the six-point functions \eqref{eq 6 point correlator Z ov Z}, \eqref{eq 6 point correlator ov Z ov Z} and \eqref{eq 6 point correlator Z Z} as well as  use the OPE
\begin{align}
\sigma^-_{\a} (x) \sigma^+_{\b} (1) \sim \left( 2 \pi \frac{\Gamma(1-\alpha)\, \Gamma(1+\alpha-\beta)\, \Gamma(\beta)}{\Gamma(\alpha)\, \Gamma(\beta -\alpha)\, \Gamma(1-\beta)}\right)^{\frac{1}{4}} (1-x)^{-\a(1-\b)} \sigma^+_{\b-\a} (1)
\end{align}
one obtains the five-point functions \eqref{eq 5 point correlator Z  ov Z}, \eqref{eq 5 point correlator Z Z} and \eqref{eq 5 point correlator ov Z  ov Z},
which were evaluated in the previous section, respectively.

Given all the six-point correlators we determine the four-point correlators containing the excited bosonic twist fields by taking various limits of the six-point correlators. Performing the limit $z\rightarrow 0, w\rightarrow x $ in  \eqref{eq 6 point correlator Z ov Z} one obtains
\begin{align}
&\langle  \tau^+_{\a} (0)  \widetilde{\tau}^-_{\a}(x) \, \sigma^+_{\b} (1)\sigma^-_{\b} (\infty)\rangle \\
& =
\sqrt{\frac{2 \pi }{I(x)}}   x^{\a(\a-3)} (1-x)^{-\a(1-\b)+\frac{\a}{2} -\frac{3 \b}{2}}
 \sum_{p,q} 
\bigg\{  - \a (1-x)^{\b-\a} \frac{B_2  H_2[1-x]\, }{I(x)}  \, _2F_1[-\a,\b,1,x]
\nn \\
& + 
\frac{ B_2 G_1[x]  }{I(x)}  \big( \a H_2[1-x] - \b \, _2F_1[-\a,\b,1-\a+\b,1-x] \big) 
  +\a (1-x)^{\b-\a}\, \frac{_2F_1[-\a,\b,1,x] }{G_1[x]}  \nn \\
& \hspace{3mm}  -\frac{\left( G_1[x] v_b +  \frac{\sin(\pi \a)}{\pi} B_1 H_1[1-x] e^{i \pi \a} v_a \right) \left( G_2[x] v_b -  \frac{\sin(\pi \a)}{\pi} B_2 H_2[1-x] e^{i \pi \a} v_a \right)}{I^2(x)} \nn
\bigg\} e^{-S_{cl}} \nn 
\end{align}
which after reinstalling all $x_i$ dependence we get:
\begin{align}
&\langle  \tau^+_{\a} (x_1)  \widetilde{\tau}^-_{\a}(x_2) \, \sigma^+_{\b} (x_3)\sigma^-_{\b} (x_4)\rangle  \label{eq corr tautilde tau}\\
& =
\sqrt{\frac{2 \pi }{I(x)}}   x_{12}^{\a(\a-3)}  x^{-\b(1-\b)}_{34}  \left(\frac{x_{14}x_{23}}{x_{13} x_{24}}\right)^{\a\b-\frac{\a}{2}-\frac{3 \b}{2}}
 \sum_{p,q} 
\bigg\{  - \a (1-x)^{\b-\a} \frac{B_2 H_2[1-x]\, }{I(x)}  \, _2F_1[-\a,\b,1,x]
\nn \\
& + 
\frac{ B_2 G_1[x]  }{I(x)}  \big( \a H_2[1-x] - \b \, _2F_1[-\a,\b,1-\a+\b,1-x] \big) 
  +\a (1-x)^{\b-\a}\, \frac{_2F_1[-\a,\b,1,x] }{G_1[x]}  \nn \\
 &  \hspace{3mm}  -\frac{\left( G_1[x] v_b +  \frac{\sin(\pi \a)}{\pi} B_1 H_1[1-x] e^{i \pi \a} v_a \right) \left( G_2[x] v_b -  \frac{\sin(\pi \a)}{\pi} B_2 H_2[1-x] e^{i \pi \a} v_a \right)}{I^2(x)} \nn
\bigg\} e^{-S_{cl}} \,\,,\nn 
\end{align}
where again $x=\frac{x_{12} x_{34}}{x_{13} x_{24}}$ and the world-sheet instanton contribution is given by \eqref{eq WS four 2 ind angles}.
The remaining four-point correlator functions for two independent angles are displayed in appendix \ref{app correlators two independent angle}.

\section{Conclusions}

In this paper we have derived the three- and four-point correlators containing exited bosonic twist fields. More specifically, we computed correlators that consist of regular bosonic twist fields $\sigma$ as well as excited and doubly excited bosonic twist fields $\tau$ and $\omega$, respectively. The knowledge of these correlators is required in the computation of lifetime and decay rates of string excitations that arise at intersections of D-branes. Assuming a low string scale scenario those stringy states can be significantly lighter than the first Regge excitations of the gauge bosons, and thus significantly lighter than the string scale and potentially observable at LHC.

In order to evaluate these correlators, we took a detour and determined five- and six-point correlators containing the conformal fields $\partial Z$ and $\partial \ov Z$ as well as the regular bosonic twist fields $\sigma$. Given those five- and six-point correlators we performed various limits to derive the three- and four-point correlators containing higher excited bosonic twist fields.

The detailed computation of amplitudes that contain stringy states localized at the intersection of two D-brane stacks as well as their decay rates and lifetime is relegated to future work. However let us already point out a few phenomenological implications of our results.

Our results reveal that the three-point correlator containing just one excited bosonic twist field $\tau$ does not contain a purely quantum part, but is dictated by the classical solution. That implies that the decay rate of the first excited stringy scalar state, localized at the intersection of D-brane stacks, into two massless fermions, depends crucially (not just via world-sheet instanton suppression) on the displacement in the compactification manifold between the stringy massive state and the massless states localized at different D-brane intersections. 

On the other hand the displacement in the compactification manifold between different D-brane intersection points is related to the observed Yukawa coupling hierarchies of the massless fermions. This potentially allows one to obtain bounds on the decay rate of light stringy states in terms of observed mass hierarchies. We now have all of the components assembled to do such an analysis. 

As already shown in \cite{Anastasopoulos:2011hj}, in $2\rightarrow 2$ processes involving chiral superfields, only the \emph{second} state in the tower is exchanged, and these processes would be suppressed by the Yukawa couplings - so the bounds on these states will be drastically weaker than those for string Regge excitations. Indeed, from dijet searches for related resonances \cite{CMS:kxa,ATLAS:2012pu,ATLAS:2012qjz}, since the cross-section is so suppressed we can infer that the bounds on such states will be much less than a TeV. This raises the intriguing prospect that the string scale could be just out of reach of the LHC, but the  light stringy states could be hiding in plain sight.

\section*{Acknowledgements}
We acknowledge M. Bianchi, M. Cveti{\v c}, F. Morales, O. Schlotterer and  S. Stieberger for interesting discussions and correspondence. 
P.~A. is supported by the Austrian Science Fund (FWF) program M 1428-N27. 
M.~D.~G. was supported by a Marie Curie Intra European Fellowship within the 7th European Community Framework Programme and by ERC advanced grant 226371.
The work of R.~R. was partly supported by the German Science Foundation (DFG) under the Collaborative Research Center (SFB) 676 ``Particles, Strings and the Early Universe".
R.~R. thanks the TU Wien for hospitality.
P.~A. thanks Ecole Polytechnique for hospitality during last stage of this work.

\newpage
\appendix

\section{OPE's of the twist fields
\label{app OPE}}

Here, we provide the OPE's of $\partial Z(x)$ and $\partial \ov Z(x)$ on twisted fields:
\bea
\begin{tabular}{llll}
&$\partial Z (z) \, \sigma^+_{\a} (w) \sim (z-w)^{\a-1} \tau^+_{\a} (w)    $&$\qquad \partial \ov Z (z) \, \sigma^+_{\a} (w) \sim (z-w)^{-\a} \widetilde{\tau}^+_{\a} (w) $\\ 
&$\partial Z (z) \, \tau^+_{\a} (w) \sim (z-w)^{\a-1} \omega^+_{\a} (w)   $&$\qquad \partial \ov Z (z) \, \tau^+_{\a} (w) \sim (z-w)^{-\a-1} \sigma^+_{\a} (w) $
\\
&$\partial Z (z) \, \omega^+_{\a} (w) \sim (z-w)^{\a-1} \rho^+_{\a} (w)   $&$\qquad \partial \ov Z (z) \, \omega^+_{\a} (w) \sim (z-w)^{-\a-1} \tau^+_{\a} (w) 
$\\
&$\partial Z (z) \, \widetilde{\tau}^+_{\a} (w) \sim (z-w)^{-2+\a} \sigma^+_{\a} (w)   $&$\qquad \partial \ov Z (z) \, \widetilde{\tau}^+_{\a} (w) \sim (z-w)^{-\a} \widetilde{\omega}^+_{\a} (w) 
$\\ 
&$\partial Z (z) \, \sigma^-_{\a} (w) \sim (z-w)^{-\a} \tau^-_{\a} (w)    $&$\qquad \partial \ov Z (z) \, \sigma^-_{\a} (w) \sim (z-w)^{\a-1} \widetilde\tau^-_{\a} (w)
$\\
&$\partial Z (z) \, \tau^-_{\a} (w) \sim (z-w)^{-\a} \omega^-_{\a} (w)  $&$\qquad \partial \ov Z (z) \, \tau^-_{\a} (w) \sim (z-w)^{-2+\a} \sigma^-_{\a} (w)
$\\
&$\partial Z (z) \, \widetilde\tau^-_{\a} (w) \sim (z-w)^{-1+\a} \sigma^-_{\a} (w)    $&$\qquad \partial \ov Z (z) \, \widetilde \tau^-_{\a} (w) \sim (z-w)^{-1-\a} \widetilde{\omega}^-_{\a} (w) $\\
&$\partial Z (z) \, \widetilde\omega^-_{\a} (w) \sim (z-w)^{-1+\a} \widetilde\tau^-_{\a} (w)    $&$\qquad \partial \ov Z (z) \, \widetilde\omega^-_{\a} (w) \sim (z-w)^{-1-\a} \widetilde{\rho}^-_{\a} (w)$
\end{tabular}~~~~
\eea

With these OPE's one can determine the conformal dimension of the respective twist fields. We summarize our findings in table \ref{table conformal dimensions}.
\begin{table}[h] \centering
\begin{tabular}{| l | l || l | l |}
\hline
Fields & conformal dimensions & Fields & conformal dimensions \\ \hline \hline
$\sigma^+_{\a}$    &   $\frac{1}{2}\a(1-\a)$   &   $\sigma^-_{\a}$    &    $\frac{1}{2} \a(1-\a) $    \\
$\tau^+_{\a}$    &   $\frac{1}{2}\a(3-\a)$   &  $\tau^-_{\a}$    &   $\frac{1}{2}( 2+\a) (1-\a)$  ~~~~~~~~~~~~    \\
$\omega^+_{\a}$    &   $\frac{1}{2}\a(5-\a)$   &      $\omega^-_{\a}$   &    $\frac{1}{2}(\a+4)(1-\a) $  \\
$\widetilde{\tau}^+_{\a}$    &   $\frac{1}{2}(\a+2)(1-\a)~~~$   &   $\widetilde{\tau}^-_{\a}$   &    $ \frac{1}{2} \a(3-\a)$   \\
$\widetilde{\omega}^+_{\a}$    &   $\frac{1}{2}(\a+4)(1-\a)~~~~~~~~~~~~$   &    $\widetilde{\omega}^-_{\a}$   &    $\frac{1}{2} \a(5-\a) $    \\
\hline
\end{tabular}\nn
\caption{\small {The conformal dimensions of bosonic twist fields.}} 
\label{table conformal dimensions}
\end{table}
The above OPE's suggest the following identifications among twist- and anti-twist fields 
\begin{align}
\sigma^{-}_{\a} (z) = \sigma^+_{1-\a} (z) \qquad \qquad \tau^-_{\a} (z) = \tau^+_{1-\a} (z) \qquad \qquad \widetilde\tau^-_{\a} (z) = \widetilde\tau^+_{1-\a} (z)
\end{align}  
which can be easily generalized to higher excited twist fields.

\section{Various integrals \label{app integrals}}
Here we display all necessary integrals needed throughout this work. 
\begin{align} \nn
&\int^x_0 d z\, \, \widetilde\omega_{\a,\b}(z) = -e^{i\pi(\a-\b)} B(1-\a,\a)  \, \, _2F_1[\b,1-\a,1,x]\\ \nn 
&\int^1_x d z\, \, \widetilde\omega_{\a,\b}(z) = e^{-i\pi \b} (1-x)^{\a-\b} B(1-\b,\a)  \, \, _2F_1[\a,1-\b,1+\a-\b,1-x]\\ \nn 
&\int^x_0 d z\, \, z\,\widetilde\omega_{\a,\b}(z) = -e^{i\pi(\a-\b)} x B(2-\a,\a)  \, \, _2F_1[\b,2-\a,2,x]\\ \nn
&\int^1_x d z\, \, z\,\widetilde\omega_{\a,\b}(z) = e^{-i\pi \b} (1-x)^{\a-\b} B(1-\b,\a)  \, \, _2F_1[-1+\a,1-\b,1+\a-\b,1-x]
\end{align}
%

\section{Vanishing of the correlators $\langle \partial Z_{qu} \sigma_\a \sigma_\b  \sigma_\c\rangle $ \label{app vanishing}}

Here we demonstrate that the quantum part of amplitudes involving a single singly-excited twist field must vanish. To do this, we simply construct the function 
\begin{align}
f(z) \equiv \frac{\langle \partial Z_{qu} (z) \prod_i \sigma_{\alpha_i} (x_i) \rangle}{\langle \prod_i \sigma_{\alpha_i} (x_i) \rangle}~,~~~~~~~~
\tilde{f}(z) \equiv \frac{\langle \partial \ov{Z}_{qu} (z)\prod_i \sigma_{\alpha_i} (x_i) \rangle}{\langle \prod_i \sigma_{\alpha_i} (x_i) \rangle}
\end{align}
where $i$ runs from $0$ to $L$. The functions $f(z)$ and  $\tilde{f}(z)$ are holomorphic functions which do exhibit the poles 
\begin{align}
f(z) \sim (z-x_i)^{\alpha_i - 1} ~,~~~~~~~~ \tilde{f}(z) \sim (z-x_i)^{-\alpha_i }.
\end{align}
in agreement \eqref{eq OPE twist fields}. Since $\partial Z, \partial \ov{Z}$ have unit conformal weight, we furthermore require $f(z), \tilde{f}(z) \sim z^{-2}$ as $z\rightarrow \infty$. 

For $L$ twist field insertions, there are $L-2$ independent contours that we can choose for boundary conditions. There are $L=1$ contours for $L$ points, but the boundary must form a closed polygon, so one is not independent. There are also and $L-2$ differentials which make up $f$ and $\tilde{f}$ in total. We construct the cut differentials
\begin{align}
\omega (z) \equiv \prod_i (z-x_i)^{\alpha_i -1}  \sim z^{M-L}~,~~~~~~~~
\tilde{\omega}(z) \equiv \prod_i (z-x_i)^{-\alpha_i} \sim z^{-M}
\end{align}
where $M \equiv \sum \alpha_i$. Then we can write 
\begin{align}
f(z) \equiv \omega(z) p(z) ~,~~~~~~~~
\tilde{f} (z) \equiv \tilde{\omega}(z) \tilde{p} (z)
\end{align} 
so the number of differentials is given by the number of independent polynomials $p, \tilde{p}$ which have order $L-M-2, M-2$, respectively. This is determined by counting the poles and allowing one parameter for the overall normalisation; the total number is thus $(L-M-2+1) + (M-2+1) = L-2$. These are forced by the boundary conditions to be real parameters. We can then solve the boundary conditions for the quantum amplitudes
\begin{align}
\int_{C_i} dZ = \int_{C_i} d\ov{Z} =& 0\,\,.
\end{align}
These are a non-degenerate set of $L-2$ equations in $L-2$ real parameters, and hence all of the correlators containing just one $\partial Z_{qu}$ are zero.

\section{Three-point correlators containing excited bosonic twist fields \label{app three point}}
From the correlator $\langle \partial Z \partial Z\sigma^+_{\alpha}\sigma^+_{\beta} \sigma^+_{\gamma} \rangle$  (see eq. \eqref{eq 5 point correlator Z Z}) we obtain
\begin{align}
\langle \tau^+_{\alpha}(x_1) \tau^+_{\beta}(x_2) \sigma^+_{\gamma}(x_3) \rangle &=-  (2\pi)^{\frac{1}{4}}
\left(\frac{\Gamma(1-\alpha)\Gamma(1-\beta)\Gamma(1-\gamma)}{\Gamma(\alpha)\Gamma(\beta)\Gamma(\gamma)} \right)^{\frac{5}{4}}
\\ & \hspace{-30mm} \times \sum_{n} \left\{ 1-\frac{\sin(\pi \alpha) \sin(\pi \beta)}{\pi \sin(\pi \gamma)} |v_a|^2 \right\}\,x^{-\alpha \beta-\alpha -\beta}_{12} x^{-\alpha(2+\gamma) -\gamma+1}_{13}
x^{-\beta(2+\gamma) -\gamma+1}_{23}\, e^{-S_{cl}} \nn
\end{align}
and 
\begin{align}
\langle \omega^+_{\alpha}(x_1) \sigma^+_{\beta}(x_2) \sigma^+_{\gamma}(x_3) \rangle &= - (2\pi)^{\frac{1}{4}}
\left(\frac{\Gamma(1-\alpha)\Gamma(1-\beta)\Gamma(1-\gamma)}{\Gamma(\alpha)\Gamma(\beta)\Gamma(\gamma)} \right)^{\frac{5}{4}}
\\ & \hspace{-30mm} \times \sum_{n} \left\{ 1-\frac{\sin(\pi \alpha) \sin(\pi \beta)}{\pi \sin(\pi \gamma)} |v_a|^2 \right\}\,x^{-\alpha(2+ \beta)}_{12} x^{-\alpha(2+\gamma) }_{13}
x^{-\beta(2+\gamma) -2\gamma+2}_{23}\, e^{-S_{cl}} \nn
\end{align}
with the world-sheet contributions given by \eqref{eq WS instanton three-point}.

\section{Four-point correlators containing excited bosonic twist fields \label{app four point}}

\subsection{Various four-point correlators with one independent angle\label{app correlators one independent angle}}

Here we display all the different four point correlators that can be derived from the six-point correlators \eqref{eq six point g}, \eqref{eq six point k} and \eqref{eq six point m}\footnote{Note that we do not display all possible limits since the missing correlators can be obtained by the identifications between boson twist and anti-twist fields.}. 
From the correlator $\langle \partial \ov Z \sigma^+_{\alpha} \sigma^+_{1-\alpha} \sigma^+_{\alpha} \sigma^+_{1-\alpha}  \rangle$ one obtains
\begin{align}
\langle  \widetilde{\tau}^+_{\alpha} (x_1) \sigma^-_{\alpha} (x_2) \sigma^+_{\alpha} (x_3) \sigma^-_{\alpha} (x_4) \rangle &=
- \left( \frac{x_{24}}{x_{12} x_{14}}\right)^{(1-\a) (1+\a)}
 \left( \frac{x_{13}}{x_{23} x_{34}}\right)^{\a (1-\a)}
\\& \hspace{-40mm} \times
{\frac{ \sin^\frac{3}{2}(\pi \alpha)}{2 \pi \,  _2F_1^\frac{1}{2}[\alpha,1-\alpha,1,x]\,    _2F^\frac{3}{2}_1[\alpha,1-\alpha,1,1-x]} } 
\sum_{p,q} \left( {v_b-e^{i\pi \alpha} v_a  \tau(x)} \right)\,e^{-S_{cl}} 
\,\,. \nn
\end{align}

From the correlator $\langle \partial Z \partial \ov Z \sigma^+_\a \sigma^-_{\a} \sigma^+_\a \sigma^-_{\a}  \rangle$ one gets
\begin{align}
&\langle  \tau^+_{\a} (x_1) \sigma^-_{\a}(x_2) \, \widetilde{\tau}^+_{\a} (x_3)\sigma^-_{\a} (x_4)\rangle \\& \hspace{15mm}= - \frac{ \a(1-\a) }{ 2} 
\frac{ \sqrt{\pi} ~ \left(x_{12} x_{14} \right)^{-\a(2-\a)}   \left( x_{23} x_{34} \right)^{-(1-\a)(1+\a)} x^{\a(1-\a)}_{13}  x^{1+\a-\a^2}_{24} }{\sqrt{\Gamma(\a)\, \Gamma(1-\a)\,_2F_1[\a,1-\a,1,x]\, _2F_1[\a,1-\a,1,1-x]}} \nn\\
&\hspace{20mm}  \times \sum_{p,q}  \left\{ x \frac{_2F_1[1-\a,\a,2,x]}{_2F_1[1-\a,\a,1,x]} - (1-x) \frac{_2F_1[1-\a,\a,2,1-x]}{_2F_1[1-\a,\a,1,1-x]} 
\right.  \nn \\
& \hspace{40mm} \left.
-\frac{\sin^2(\pi \a)}{ 2 \pi^2 \a(1-\a)}\frac{v^2_b - e^{2 i \pi \a} v^2_a \tau^2(x)}{_2F_1[1-\a,\a,1,1-x]^2}
 \right\}  e^{-S_{cl}} \nn\,\,.
\end{align}
From the correlator $\langle  \partial \ov Z \partial \ov Z \sigma^+_\a \sigma^-_{\a} \sigma^+_\a \sigma^-_{\a}  \rangle$ one obtains
\begin{align}
& \langle  \widetilde{\omega}^+_{\a} (x_1) \sigma^-_{\a}(x_2) \, \sigma^+_{\a} (x_3)\sigma^-_{\a} (x_4)\rangle 
\\& \hspace{15mm} =\frac{\a e^{2i \pi \a}}{2} \frac{ \sqrt{\pi} ~  \left(\frac{x_{24}}{ x_{12} x_{14}} \right)^{(2+\a)(1-\a)} \left( \frac{x_{13}} { x_{23} x_{34}}\right)^{\a(1-\a)}  }{\sqrt{\Gamma(\a) \Gamma(1-\a)\,_2F_1[\a,1-\a,1,x]\, _2F_1[\a,1-\a,1,1-x]}} \nn\\  
&\hspace{20mm} \times \sum_{p,q}  \bigg\{ 1-  \frac{_2F_1[-\a,\a,1,1-x]}{_2F_1[1-\a,\a,1,1-x]} -  \frac{_2F_1[-\a,\a,1,x]}{_2F_1[1-\a,\a,1,x]}   \nn \\
& \hspace{40mm}
 +\frac{ \sin^2(\pi \a)} {2 \pi^2 \a} \frac{\left( v_b - e^{i\pi \a} v_a \tau(x)\right)^2  } {_2F_1[\a,1-\a,1,1-x]^{2} }   
 \bigg\}  e^{-S_{cl}} 
\nn \end{align} 
\begin{align}
& \langle  \widetilde{\tau}^+_\a (x_1) \widetilde{\tau}^-_{\a}(x_2) \, \sigma^+_{\a} (x_3)\sigma^-_{\a} (x_4)\rangle \\& \hspace{15mm}= -\frac{\a e^{2i \pi \a} }{2 } \frac{ \sqrt{\pi} ~  x^{-\a(1-\a)-1}_{12} \left(\frac{x_{13}}{x_{23}}\right)^{\a(2-\a)} \left(\frac{x_{24}}{x_{14}}\right)^{(1+\a)(1-\a)} x^{-\a(1-\a)}_{34} }{\sqrt{\Gamma(\a) \Gamma(1-\a) \, _2F_1[\a,1-\a,1,x]\, _2F_1[\a,1-\a,1,1-x]}} \nn
 \\  
&\hspace{20mm} \times \sum_{p,q} 
 \bigg\{ 1-  \frac{_2F_1[-\a,\a,1,1-x]}{_2F_1[1-\a,\a,1,1-x]} -  \frac{_2F_1[-\a,\a,1,x]}{_2F_1[1-\a,\a,1,x]} \nn \\  \nn
& \hspace{40mm}
 +\frac{\sin^2(\pi \a)} {2 \pi^2 \a} \frac{\left( v_b - e^{i\pi \a} v_a \tau(x)\right)^2  } {_2F_1[\a,1-\a,1,1-x]^{2} }   
 \bigg\}  e^{-S_{cl}} 
\nn 
\end{align} 
\begin{align}
& \langle  \widetilde{\tau}^+_\a (x_1)\sigma^-_{\a}(x_2) \widetilde{\tau}^+_\a(x_3)\sigma^-_{\a} (x_4)\rangle \\  &\hspace{20mm} = 
\frac{\a e^{2\pi i \a}}{2 } \frac{ \sqrt{\pi} ~  \left( x_{12} x_{14} x_{23} x_{34}\right)^{-(\a+1)(1-\a)} x^{\a(1-\a)}_{13} x^{(1-\a)(\a+2)}_{24} }{\sqrt{\Gamma(\a) \Gamma(1-\a)\, _2F_1[\a,1-\a,1,x]\, _2F_1[\a,1-\a,1,1-x]}} \nn\\  
&\hspace{35mm} \times \sum_{p,q}  \bigg\{ 1-  \frac{_2F_1[-\a,\a,1,1-x]}{_2F_1[1-\a,\a,1,1-x]} -  \frac{_2F_1[-\a,\a,1,x]}{_2F_1[1-\a,\a,1,x]}  \nn \\
& \hspace{50mm}
 +\frac{ \sin^2(\pi \a)} {2 \pi^2 \a} \frac{\left( v_b - e^{i\pi \a} v_a \tau(x)\right)^2  } {_2F_1[\a,1-\a,1,1-x]^{2} }   
 \bigg\}  e^{-S_{cl}} 
\nn
 \end{align} 
while from the correlator $\langle \partial Z \partial Z  \sigma^+_{\a} \sigma^-_{\a}  \sigma^+_{\a} \sigma^-_{\a} \rangle $ one gets
\begin{align}
&\langle  \omega^+_\a (x_1)\sigma^-_{\a}(x_2) \,\sigma^+_\a(x_3)\sigma^-_{\a} (x_4)\rangle 
\\ \nn & \hspace{15mm}=
\frac{(1-\a) e^{-2i \pi \a}}{2 } \frac{ \sqrt{\pi} ~ x^{-\a(3-\a)}_{12}      \left(\frac{x_{13}}{x_{23}}\right)^{\a(1-\a)}    \left(\frac{x_{24}}{x_{14}}\right)^{\a(3-\a)}     x^{-\a(1-\a)}_{34}}{\sqrt{\Gamma(\a) \Gamma(1-\a)\,_2F_1[\a,1-\a,1,x]\, _2F_1[\a,1-\a,1,1-x]}} \\ 
&\hspace{25mm} \times \sum_{p,q} 
\bigg\{1 - x \frac{ _2F_1[2-\a,\a,1,1-x] }{ _2F_1[1-\a,\a,1,1-x]} -(1-x) 
\frac{ _2F_1[2-\a,\a,1,x] }{ _2F_1[1-\a,\a,1,x]}  \nn \\  \nn
& \hspace{40mm}
 +\frac{ \sin^2(\pi \a)} {2 \pi^2 (1-\a)} \frac{\left( v_b + e^{i\pi \a} v_a  \tau(x)\right)^2  } {_2F_1[\a,1-\a,1,1-x]^{2} }   
 \bigg\}  e^{-S_{cl}} 
\end{align}
\begin{align}
& \langle  \tau^+_\a (x_1) \tau^-_\a (x_2) \,\sigma^+_\a(x_3)\sigma^-_{\a} (x_4)\rangle  \\ \nn & \hspace{10mm}=
\frac{(1-\a) e^{-2\pi i \a }}{2 } \frac{ \sqrt{\pi} ~  x^{-\a(1-\a)-1}_{12} \left( \frac{x_{13}}{x_{23}}\right)^{(1+\a)(1-\a)} \left( \frac{x_{24}}{x_{14}}\right)^{\a(2-\a)} x^{-\a(1-\a)}_{34}}{\sqrt{\Gamma(\a) \Gamma(1-\a)\,_2F_1[\a,1-\a,1,x]\, _2F_1[\a,1-\a,1,1-x]}} \nn \\ 
&\hspace{25mm} \times \sum_{p,q} 
\bigg\{1 - x \frac{ _2F_1[2-\a,\a,1,1-x] }{ _2F_1[1-\a,\a,1,1-x]} -(1-x) 
\frac{ _2F_1[2-\a,\a,1,x] }{ _2F_1[1-\a,\a,1,x]}  \nn \\  \nn
& \hspace{40mm}
 +\frac{ \sin^2(\pi \a)} {2 \pi^2 (1-\a)} \frac{\left( v_b + e^{i\pi \a} v_a \tau(x)\right)^2  } {_2F_1[\a,1-\a,1,1-x]^{2} }   
 \bigg\}  e^{-S_{cl}} 
\end{align}
\begin{align}
&\langle  \tau^+_{\a} (x_1) \sigma^-_{\a}(x_2) \,\tau^+_{\a} (x_3)\sigma^-_{\a} (x_4)\rangle \\  & \hspace{15mm}=
 \frac{(1-\a)  e^{-2\pi i \a }  }{2 } \frac{ \sqrt{\pi} ~ 
\left( x_{12} x_{14} x_{23} x_{24}\right)^{-\a(2-\a)}   x^{\a(1-\a)}_{13}  x^{\a(3-\a)}_{24}  }{\sqrt{\Gamma(\a) \Gamma(1-\a)\,_2F_1[\a,1-\a,1,x]\, _2F_1[\a,1-\a,1,1-x]}} \nn \\ 
&\hspace{30mm} \times \sum_{p,q} 
\bigg\{1 - x \frac{ _2F_1[2-\a,\a,1,1-x] }{ _2F_1[1-\a,\a,1,1-x]} -(1-x) 
\frac{ _2F_1[2-\a,\a,1,x] }{ _2F_1[1-\a,\a,1,x]}  \nn \\  \nn
& \hspace{50mm} 
 +\frac{\sin^2(\pi \a)} {2 \pi^2 (1-\a)} \frac{\left( v_b + e^{i\pi \a} v_a \tau(x)\right)^2  } {_2F_1[\a,1-\a,1,1-x]^{2} }   
 \bigg\}  e^{-S_{cl}} 
\end{align}

\subsection{Various four-point correlators with two independent angle
\label{app correlators two independent angle}}

Here we display all the different four point correlators that can be derived from the five point correlator \eqref{eq ov dZ ssss ab} and the six-point correlators , \eqref{eq six point g ab}, \eqref{eq six point k ab} and \eqref{eq six point m ab}.

From the five-point correlator $\langle \partial \ov Z \sigma^+_{\a}  \sigma^-_{\a} \, \sigma^+_{\b}\sigma^-_{\b}\rangle $ we obtain the four-point correlator
\begin{align}
&\langle \tilde \tau^+_{\alpha} (x_1) \sigma^-_{\alpha} (x_2) \sigma^+_{\b} (x_3) \sigma^-_{\b} (x_4) \rangle
\label{eq ov tildetsss ab all x1x2x3x4}\\
& \hspace{20mm} = 
\frac{ \sqrt{ 2 \pi} ~ 
x_{12}^{-1+\a^2}
\left(
\frac{x_{13}}{x_{23}}
\right)^{-\a\b +\frac{\a}{2} +\frac{\b}{2}}
\left(
\frac{x_{14}}{x_{24}}
\right)^{\a\b +\frac{\a}{2} -\frac{\b}{2}-1}
x_{34}^{(\b-1)\b}
}{\sqrt{I(x)}}   
\nn
\\
&  \hspace{40mm} \times \sum_{p,q} 
\frac{\left( G_1[x] v_b -  \frac{\sin(\pi \a)}{\pi} (1-x)^{\b-\a} B_2 H_2[1-x] e^{i \pi \a} v_a \right)}{I(x)} \nn
\end{align}

From the  six-point correlator $\langle \partial Z \partial \ov Z \sigma^+_{\a}  \sigma^-_{\a} \, \sigma^+_{\b}\sigma^-_{\b}\rangle $ we obtain the four-point correlators 
\begin{align}
&\langle  \tau^+_{\a} (x_1) \sigma^-_{\a}(x_2) \, \widetilde{\tau}^+_{\b} (x_3)\sigma^-_{\b} (x_4)\rangle \\
& =
\sqrt{\frac{2 \pi }{I(x)}} e^{i \pi (\b-\a)}   x_{12}^{-\a(2-\a)}   x^{-\a\b -\frac{\a}{2} +\frac{3\b}{2} }_{13} 
x^{\a\b -\frac{\a}{2} -\frac{3\b}{2} }_{14} 
x^{\a\b +\frac{\a}{2} -\frac{\b}{2} -1}_{23}  x^{-\a\b+\frac{\a}{2} +\frac{\b}{2} +1}_{24}
x^{-(1+\b)(1-\b)}_{34} 
\nn \\
& 
\nn  \times \sum_{p,q} \bigg\{ - a(1-x) +  \a\, \frac{ B_2 G_1[x]  H_2[1-x]\, }{I(x)} - a (1-x)^{\b-\a} \frac{B_2  H_2[1-x]\, }{I(x)}  \, _2F_1[-\a,\b,1,x]
 \\&
\hspace{6mm} - b\,  \frac{B_2 G_1[x] }{I(x)}  \, _2F_1[-\a,\b,1-\a+\b,1-x] +\a (1-x)^{\b-\a}\, \frac{_2F_1[-\a,\b,1,x] }{G_1[x]} \nn \\
 & \hspace{3mm}  - \frac{\left( G_1[x] v_b +  \frac{sin(\pi \a)}{\pi} B_1 H_1[1-x] e^{i \pi \a} v_a \right) \left( G_2[x] v_b -  \frac{sin(\pi \a)}{\pi} B_2 H_2[1-x] e^{i \pi \a} v_a \right)}{I^2(x)} \nn
\bigg\} e^{-S_{cl}} \nn 
\end{align}
while from the six-point correlator $\langle \partial \ov  Z \partial \ov Z \sigma^+_{\a}  \sigma^-_{\a} \, \sigma^+_{\b}\sigma^-_{\b}\rangle $ we get the four-point correlators
\begin{align}
& \langle  \widetilde{\omega}^+_{\a} (x_1) \sigma^-_{\a}(x_2) \, \sigma^+_{\b} (x_3)\sigma^-_{\b} (x_4)\rangle
\\
&=  
\sqrt{ \frac{2 \pi }{I(x)}}   x^{-(\a+2)(1-\a)}_{12} x^{-\a\b+\frac{3\a}{2}-\frac{\b}{2}}_{13} x^{\a\b+\frac{\a}{2} +\frac{\b}{2}-2}_{14} x^{\a\b-\frac{3\a}{2} +\frac{\b}{2}}_{23} x^{-\a\b-\frac{\a}{2} -\frac{\b}{2}+2}_{24} x^{-\b(1-\b)}_{34}\nn \\
\nn
& \hspace{1mm} \times \sum_{p,q} \bigg\{  
\frac{B_2}{I(x)} \Big( {\a}~ {G_1[x] H_2[1-x]} 
- {\b}~ {G_1[x]} \, _2F_1[-\a,\b,1-\a+\b,1-x] \nn \\
& \left. \hspace{30mm} - {\a}~ (1-x)^{\b-\a} {H_2[1-x]} \, _2F_1[-\a,\b,1,x] \Big) \nn  \right. \\
& \hspace{40mm}
+\frac{\left( G_1[x] v_b -  \frac{\sin(\pi \a)}{\pi} (1-x)^{\b-\a} B_2 H_2[1-x] e^{i \pi \a} v_a \right)^2}{I^2(x)} 
\bigg\}e^{-S_{cl}} \nn
\end{align} 
\begin{align}
& \langle  \widetilde{\tau}^+_\a (x_1) \widetilde{\tau}^-_{\a}(x_2) \, \sigma^+_{\b} (x_3)\sigma^-_{\b} (x_4)\rangle \\
&= 
- 
  \sqrt{ \frac{2 \pi }{I(x)}} x^{-\a(1-\a)-1}_{12}   x^{-\a\b+\frac{3\a}{2}+\frac{\b}{2}}_{13} x^{\a\b+\frac{\a}{2} -\frac{\b}{2}-1}_{14} x^{\a\b-\frac{3\a}{2} -\frac{\b}{2}}_{23} x^{-\a\b-\frac{\a}{2} +\frac{\b}{2}+1}_{24} x^{-\b(1-\b)}_{34}\nn\\
\nn
& \hspace{1mm}  \times \sum_{p,q}  \bigg\{
\frac{B_2}{I(x)} \Big( {\a}~ {G_1[x] H_2[1-x]}  - {\b}~ {G_1[x]} \, _2F_1[-\a,\b,1-\a+\b,1-x] \nn\\ 
& \left.\hspace{30mm} - {\a}~ (1-x)^{\b-\a} {H_2[1-x]} \, _2F_1[-\a,\b,1,x] \Big)\nn  \right. \\
& \hspace{40mm}
+\frac{\left( G_1[x] v_b -  \frac{\sin(\pi \a)}{\pi} (1-x)^{\b-\a} B_2 H_2[1-x] e^{i \pi \a} v_a \right)^2}{I^2(x)} \bigg\} e^{-S_{cl}}
\nn
\end{align} 
%
%
\begin{align}
& \langle  \widetilde{\tau}^+_\a (x_1)\sigma^-_{\a}(x_2) \widetilde{\tau}^+_\b(x_3)\sigma^-_{\b} (x_4)\rangle  \\
&=   
  \sqrt{ \frac{2 \pi }{I(x)}} 
x^{-(\a+1)(1-\a)}_{12}   x^{-\a\b+\frac{\a}{2}+\frac{\b}{2}}_{13} x^{\a\b+\frac{\a}{2} -\frac{\b}{2}-1}_{14} x^{\a\b-\frac{\a}{2} +\frac{\b}{2}-1}_{23} x^{-\a\b-\frac{\a}{2} -\frac{\b}{2}+2}_{24} x^{-(\b+1)(1-\b)}_{34}\nn \\
\nn
& \hspace{1mm}  \times \sum_{p,q} \bigg\{
\frac{B_2}{I(x)} \Big({\a}~ {G_1[x] H_2[1-x]}  - {\b}~ {G_1[x]} \, _2F_1[-\a,\b,1-\a+\b,1-x]  \nn\\ 
& \hspace{30mm} - {\a}~ (1-x)^{\b-\a} {H_2[1-x]} \, _2F_1[-\a,\b,1,x] \Big)\nn
\nn \\
& \hspace{40mm}
+\frac{\left( G_1[x] v_b -  \frac{\sin(\pi \a)}{\pi} (1-x)^{\b-\a} B_2 H_2[1-x] e^{i \pi \a} v_a \right)^2}{I^2(x)} \bigg\} e^{-S_{cl}}\,\,. \nn
\end{align} 
From the six-point correlator  $\langle \partial Z \partial  Z   \sigma^+_{\a} \sigma^-_{\a}  \,  \sigma^+_{\b} \sigma^-_{\b} \rangle $ one can derive the following four-point correlators
\begin{align}\label{eq correlator omega +}
&\langle \omega^+_\a (x_1)\sigma^-_{\a}(x_2) \,\sigma^+_\b(x_3)\sigma^-_{\b} (x_4)\rangle  \\
&= 
 \sqrt{ \frac{2 \pi }{I(x)}}  x^{-\a(3-\a)}_{12}   x^{-\a\b-\frac{\a}{2}+\frac{3\b}{2}}_{13} x^{\a\b-\frac{3\a}{2} -\frac{3\b}{2}}_{14} x^{\a\b+\frac{\a}{2} -\frac{3\b}{2}}_{23} x^{-\a\b+\frac{3\a}{2} +\frac{3\b}{2}}_{24} x^{-\b(1-\b)}_{34}\nn \\
%
 & \hspace{10mm}  \times \sum_{p,q} 
\bigg\{
 (1-\a)~ \frac{B_1}{I(x)} \Big\{
 G_2[x] H_1[1-x] -(1-x)  H_1[1-x]\, _2F_1[2-\a,\b,1,x] \nn\\
& \left. \hspace{40mm}   -\frac{1-\b}{1-\a} x G_2[x] \, _2F_1[2-\b,\a,1+\a-\b,1-x]\Big\}\right. \nn\\
& \hspace{45mm} 
+\frac{\left( G_2[x] v_b
 +  \frac{\sin(\pi \a)}{\pi} (1-x)^{\a-\b} B_1 H_1[1-x] e^{i \pi \a} v_a \right)^2}{I^2(x)} \bigg\} e^{-S_{cl}}
 \nn
\end{align}
\begin{align}
& \langle  \tau^+_\a (x_1) \tau^-_\a (x_2) \,\sigma^+_\b(x_3)\sigma^-_{\b} (x_4)\rangle
  \\
&=  -
\sqrt{ \frac{2 \pi }{I(x)}}  x^{-\a(1-\a)-1}_{12}   x^{-\a\b-\frac{\a}{2}+\frac{\b}{2}+1}_{13} x^{\a\b-\frac{3\a}{2}-\frac{\b}{2}}_{14} 
x^{\a\b+\frac{\a}{2} -\frac{\b}{2}-1}_{23} x^{-\a\b+\frac{3\a}{2} +\frac{\b}{2}}_{24} x^{-\b(1-\b)}_{34}\nn\\
%
 & \hspace{10mm}  \times \sum_{p,q} 
\bigg\{  (1-\a)~ \frac{B_1}{I(x)} \Big(
 G_2[x] H_1[1-x] -(1-x)  H_1[1-x]\, _2F_1[2-\a,\b,1,x] \nn \\
& \left. \hspace{40mm}   -\frac{1-\b}{1-\a} x G_2[x] \, _2F_1[2-\b,\a,1+\a-\b,1-x]\Big\}
\right. \nn\\
& 
\hspace{45mm}  +\frac{\left( G_2[x] v_b
 +  \frac{\sin(\pi \a)}{\pi} (1-x)^{\a-\b} B_1 H_1[1-x] e^{i \pi \a} v_a \right)^2}{I^2(x)} \bigg\} e^{-S_{cl}}
 \nn
\end{align}
\begin{align}
& \langle  \tau^+_\a (x_1) \sigma^-_{\a} (x_2) \,\tau^+_\b(x_3)\sigma^-_{\b} (x_4)\rangle 
  \\
&=  e^{- i \pi ( \a+\b)}
\sqrt{ \frac{2 \pi }{I(x)}} x^{-\a(2-\a)}_{12}   x^{-\a\b+\frac{\a}{2}+\frac{\b}{2}}_{13} x^{\a\b-\frac{3\a}{2}-\frac{\b}{2}}_{14} 
x^{\a\b-\frac{\a}{2} -\frac{3\b}{2}}_{23} x^{-\a\b+\frac{3\a}{2} +\frac{3\b}{2}}_{24} x^{-\b(2-\b)}_{34}\nn \\
 & \hspace{10mm}  \times \sum_{p_a,p_b}  \bigg\{ 
(1-\a)~ \frac{B_1}{I(x)} \Big\{
 G_2[x] H_1[1-x] -(1-x)  H_1[1-x]\, _2F_1[2-\a,\b,1,x] \nn \\
& \hspace{40mm}   -\frac{1-\b}{1-\a} x G_2[x] \, _2F_1[2-\b,\a,1+\a-\b,1-x]\Big\}  \nn\\
& \hspace{45mm}  +\frac{\left( G_2[x] v_b
 +  \frac{\sin(\pi \a)}{\pi} (1-x)^{\a-\b} B_1 H_1[1-x] e^{i \pi \a} v_a \right)^2}{I^2(x)} \bigg\} e^{-S_{cl}}\,\,.
 \nn
\end{align}
%
%
%

\clearpage \nocite{*}

\bibliographystyle{JHEP}
\bibliography{refs.bib}

\end{document}